\documentclass[12pt]{article}

\usepackage[utf8]{inputenc}
\usepackage{textcomp}
\usepackage{amsmath}
\usepackage[shortlabels]{enumitem}
\usepackage{chetnew}
\usepackage{tikz}
\usepackage{amssymb,amsfonts,mathrsfs,dsfont,yfonts,bbm}\usetikzlibrary{calc}
\usepackage{xcolor}
\usepackage{braket}
\usepackage[normalem]{ulem}
\usepackage{import}
\usepackage{booktabs}
\usepackage{varioref}
\usepackage{bm}
\usetikzlibrary{arrows,decorations.markings}
\usetikzlibrary{decorations.pathreplacing}
\tikzset{
  ->-/.style={decoration={markings, mark=at position 0.2 with {\arrow{stealth}}},
              postaction={decorate}},
}

\usepackage{tikz-cd}
\usetikzlibrary{decorations.pathmorphing}
\newcommand{\zz}{\mathbb{Z}}

\newcommand{\p}{\partial}

\newcommand{\beqn}{\begin{eqnarray}}
\newcommand{\eeqn}{\end{eqnarray}}

  \newcommand{\PBK}[1]{\ensuremath{\begin{pmatrix}#1\end{pmatrix}}}

\newcommand{\bZ}{{\mathbb Z}}

\newcommand{\be}{\begin{equation}}
\newcommand{\ee}{\end{equation}}
\newcommand{\bea}{\begin{eqnarray}}
\newcommand{\eea}{\end{eqnarray}}

\newcommand{\CL}{\mathcal{L}}

\newcommand{\CF}{\mathcal{F}}
\newcommand{\CA}{\mathcal{A}}
\newcommand{\CN}{\mathcal{N}}

\newcommand{\CG}{\mathcal G}

\newcommand{\CD}{\mathcal{D}}
\newcommand{\CE}{\mathcal{E}}

\newcommand{\CB}{\mathcal{B}}
\newcommand{\CC}{\mathcal{C}}
\newcommand{\CZ}{\mathcal{Z}}
\newcommand{\CO}{\mathcal{O}}

\newcommand{\CT}{\mathcal{T}}
\newcommand{\CP}{\mathcal{P}}

\newcommand{\CS}{\mathcal{S}}

\newcommand{\da}{\dot{a}}
\newcommand{\db}{\dot{b}}
\newcommand{\dc}{\dot{c}}
\newcommand{\dd}{\dot{d}}
\newcommand{\de}{\dot{e}}
\newcommand{\ba}{\begin{aligned}}
\newcommand{\ea}{\end{aligned}}

\def\tpsi{\tilde \psi}

\newcommand*{\boxcoloro}{orange}
\makeatletter
\newcommand{\boxedo}[1]{\textcolor{\boxcoloro}{%
\tikz[baseline={([yshift=-1ex]current bounding box.center)}] \node [rectangle, minimum width=1ex,rounded corners,draw] {\normalcolor\m@th$\displaystyle#1$};}}
\makeatother
\newcommand*{\boxcolorr}{red}
\makeatletter
\newcommand{\boxedr}[1]{\textcolor{\boxcolorr}{%
\tikz[baseline={([yshift=-1ex]current bounding box.center)}] \node [rectangle, minimum width=1ex,rounded corners,draw] {\normalcolor\m@th$\displaystyle#1$};}}
\makeatother
\newcommand*{\boxcolorb}{blue}
\makeatletter
\newcommand{\boxedb}[1]{\textcolor{\boxcolorb}{%
\tikz[baseline={([yshift=-1ex]current bounding box.center)}] \node [rectangle, minimum width=1ex,rounded corners,draw] {\normalcolor\m@th$\displaystyle#1$};}}
\makeatother
\newcommand*{\boxcolorg}{green}
\makeatletter
\newcommand{\boxedg}[1]{\textcolor{\boxcolorg}{%
\tikz[baseline={([yshift=-1ex]current bounding box.center)}] \node [rectangle, minimum width=1ex,rounded corners,draw] {\normalcolor\m@th$\displaystyle#1$};}}
\makeatother
\newcommand*{\boxcolorp}{purple}
\makeatletter
\newcommand{\boxedp}[1]{\textcolor{\boxcolorp}{%
\tikz[baseline={([yshift=-1ex]current bounding box.center)}] \node [rectangle, minimum width=1ex,rounded corners,draw] {\normalcolor\m@th$\displaystyle#1$};}}
\makeatother
\newcommand*{\boxcolorc}{cyan}
\makeatletter
\newcommand{\boxedc}[1]{\textcolor{\boxcolorc}{%
\tikz[baseline={([yshift=-1ex]current bounding box.center)}] \node [rectangle, minimum width=1ex,rounded corners,draw] {\normalcolor\m@th$\displaystyle#1$};}}
\makeatother
\newcommand*{\boxcolory}{yellow}
\makeatletter
\newcommand{\boxedy}[1]{\textcolor{\boxcolory}{%
\tikz[baseline={([yshift=-1ex]current bounding box.center)}] \node [rectangle, minimum width=1ex,rounded corners,draw] {\normalcolor\m@th$\displaystyle#1$};}}
\makeatother

\usepackage{amsmath}	
\title{3d $\CN=4$ Mirror Symmetry, TQFTs, \\[1mm] and 't Hooft Anomaly Matching}

\author{Mahesh~K.~N.~Balasubramanian,$^{\between}$ Anindya~Banerjee,$^{\between,\varkappa}$ Matthew~Buican,$^{\between}$ Zhihao~Duan,$^{\between}$ Andrea~E.~V.~Ferrari,$^{\boxtimes,\omega}$ and Hongliang~Jiang$^{\theta}$}

\date{December 2024}
\affiliation{$^{\between}$\smallskip CTP and Department of Physics and Astronomy\\
Queen Mary University of London, London E1 4NS, UK\\[1mm]$^{\varkappa}$ University of Cincinnati, Physics Department, Cincinnati OH 45221, USA \\[1mm]$^{\boxtimes}$ Deutsches Elektronen-Synchrotron DESY, Notkestraße 85, 22607 Hamburg, Germany\\[1mm] $^{\omega}$ School of Mathematics, The University of Edinburgh, Edinburgh EH9 3FD, UK \\[1mm]$^{\theta}$ Blackett Laboratory, Imperial College, London SW7 2AZ, UK}

\abstract{Any local unitary 3d $\CN=4$ superconformal field theory (SCFT) has a corresponding \lq\lq universal" relevant deformation that takes it to a gapped phase. This deformation preserves all continuous internal symmetries, $\CS$, and therefore also preserves any 't Hooft anomalies supported purely in $\CS$. We describe the resulting phase diagram in the case of SCFTs that arise as the endpoints of renormalization group flows from 3d $\CN=4$ Abelian gauge theories with any number of $U(1)$ gauge group factors and arbitrary integer charges for the matter fields. We argue that the universal deformations take these QFTs to Abelian fractional quantum Hall states in the infrared (IR), and we explain how to match 't Hooft anomalies between the non-topological ultraviolet theories and the IR topological quantum field theories (TQFTs). Along the way, we give a proof that 3d $\CN=4$ mirror symmetry of our Abelian gauge theories descends to a duality of these TQFTs. Finally, using our anomaly matching discussion, we describe how to connect, via the renormalization group, abstract local unitary 3d $\CN=4$ SCFTs with certain 't Hooft anomalies for their internal symmetries to IR phases (partially) described by Abelian spin Chern-Simons theories.
}

\begin{document}
\setcounter{tocdepth}{2}
\maketitle
\toc
\newsec{Introduction}
The genesis of this paper was \href{https://www.london-tqft.co.uk/#24}{a journal club} one of us gave which then led to a series of questions.\footnote{The talk was partly based on work to appear~\cite{FerrariToAppear}.} Our starting point is the fact that many of the simplest 3d $\CN=4$ SUSY gauge theories have 't Hooft anomalies for their continuous flavor symmetries \cite{Bhardwaj:2022dyt,Bhardwaj:2023zix}. By anomaly matching, the same holds for their infrared (IR) fixed point superconformal field theories (SCFTs). Another important fact is that all local unitary 3d $\CN=4$ SCFTs can be gapped by turning on the following type of relevant SUSY-preserving deformation \cite{Cordova:2016xhm}
\begin{equation}\label{univDef}
\delta S\sim\int d^3x \ m\ \varepsilon^{\alpha\beta}\varepsilon_{ab}\varepsilon_{\dot a\dot b}Q^{a\dot a}_{\alpha}Q^{b\dot b}_{\beta}J+\CO(m^2)~,
\end{equation}
where $J$ is the scaling-dimension-one primary of the multiplet that contains the traceless EM tensor, $T_{\mu\nu}$, and other currents related to it by supersymmetry.\footnote{Given its universality in local theories, \eqref{univDef} is referred to as the \lq\lq universal" mass deformation in \cite{Cordova:2016xhm} .} After turning on the deformation in \eqref{univDef}, we may flow to a non-trivial TQFT. However, unlike the TQFTs one associates with twisted SUSY theories, the TQFTs produced by \eqref{univDef} are unitary\footnote{Here we are assuming the starting SCFT is unitary; if the starting SCFT is non-unitary, the IR theory may potentially be gapless.} and often semi-simple. Hence, they are typically finite (i.e., built from a finite number of simple Wilson lines satisfying finite fusion rules) and rigid (they do not admit any continuous deformations). In fact, we will argue below that this picture is generally what we expect from the $F$-theorem (provided the UV starting point has finite $F$). In the case of SCFTs arising from theories with 't Hooft anomalies for their continuous flavor symmetries, the deformation in \eqref{univDef} must preserve these symmetries (the stress tensor commutes with all continuous flavor symmetries\footnote{The deformation in \eqref{univDef} also preserves the $\mathfrak{so}(4)_R\cong \mathfrak{su}(2)_C\oplus \mathfrak{su}(2)_H$ $R$ symmetry of the theory. \label{RsymmPres}}) and 't Hooft anomalies. These observations immediately lead to the following question:

\begin{figure}
\centering

\tikzset{every picture/.style={line width=0.75pt}} 

\begin{tikzpicture}[x=0.75pt,y=0.75pt,yscale=-1,xscale=1]

\draw  [line width=1.5]  (241.11,106.36) -- (260.47,166.1) -- (176.68,152.73) -- (157.33,93) -- cycle ;
\draw  [line width=1.5]  (241.53,105.93) -- (295.86,49.51) -- (315.04,110.42) -- (260.71,166.84) -- cycle ;
\draw  [dash pattern={on 0.84pt off 2.51pt}]  (241.53,105.93) -- (241.53,165) ;
\draw [line width=1.5]    (241.53,165) -- (241.53,221) ;
\draw [line width=1.5]    (260.47,166.1) -- (260.47,243) ;
\draw [line width=1.5]    (241.53,221) -- (260.47,243) ;
\draw [color={rgb, 255:red, 208; green, 2; blue, 27 }  ,draw opacity=1 ][line width=2.25]    (241.53,105.93) -- (260.71,166.84) ;

\draw (173,102) node [anchor=north west][inner sep=0.75pt]   [align=left] {g};
\draw (285,68) node [anchor=north west][inner sep=0.75pt]   [align=left] {h};
\draw (266.53,231) node [anchor=north west][inner sep=0.75pt]   [align=left] {gh};
\draw (268,163) node [anchor=north west][inner sep=0.75pt]   [align=left] {$\displaystyle \nu $};

\end{tikzpicture}
\caption{A junction of continuous zero-form symmetry defects (corresponding to group elements, $g$, $h$, and $gh$) descended from the UV gauge theory / SCFT with an insertion of a line $v$ (in red) from the IR TQFT leads to symmetry fractionalization and a non-trivial action of the symmetry on the IR lines.}
\label{FracFig}
\end{figure}
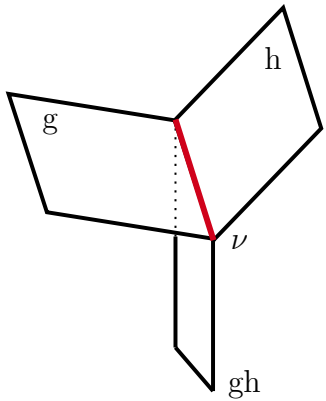

\begin{enumerate}
\item[{\bf Q1:}] How do rigid and finite TQFTs resulting from \eqref{univDef} know about the continuous SCFT flavor symmetries and their 't Hooft anomalies? In particular, how does anomaly matching work in the TQFT?
\end{enumerate}

Of particular interest to us is the phase diagram of these SCFTs and their deformations, as well as the interplay of these deformations with infrared dualities\footnote{See~\cite{Gomis:2017ixy} for a study of closely related questions in cases with less supersymmetry.} and 3d $\mathcal{N}=4$ mirror symmetry \cite{Intriligator:1996ex} in particular. Since it is quite difficult to deform strongly coupled 3d $\mathcal{N}=4$ SCFTs directly, it is convenient to first ask whether an ancestor deformation can be defined in the case of theories with UV Lagrangian descriptions:\footnote{Such descriptions might, for sufficiently rich Lagrangians and RG flows, universally be available in 3d \cite{Buican:2023efi}.}
\begin{enumerate}
\item[{\bf Q2:}] In the cases where the SCFTs in question arise via RG flows from UV Lagrangians, can we make sense of an ancestor of \eqref{univDef} in the non-conformal UV?
\end{enumerate}
As a final motivation for our present work, we note that semi-simple but non-unitary TQFTs obtained from the topological twists of two 3d $\mathcal{N}=4$ mirror families of rank-0 theories have recently been discovered to be related to level-rank dualities~\cite{Ferrari:2023fez,Creutzig:2024ljv}, and it is therefore natural to ask whether similar phenomena arise for the TQFTs obtained from the deformation in \eqref{univDef}:
\begin{enumerate}
\item[{\bf Q3:}] Can we write down the phase diagrams for 3d $\CN=4$ SCFTs deformed by \eqref{univDef} (including the ancestor deformation)? More modestly, can we write down these phase diagrams in the case of SCFTs arising from 3d $\CN=4$ Abelian gauge theories using mirror symmetry~\cite{Intriligator:1996ex,deBoer:1996ck} as a guide?
\end{enumerate}

\bigskip 

\begin{figure}
\centering

\tikzset{every picture/.style={line width=0.75pt}} 

\begin{tikzpicture}[x=0.75pt,y=0.75pt,yscale=-1,xscale=1]

\draw    (95,290) -- (524.33,290) ;
\draw [shift={(527.33,290)}, rotate = 180] [fill={rgb, 255:red, 0; green, 0; blue, 0 }  ][line width=0.08]  [draw opacity=0] (8.93,-4.29) -- (0,0) -- (8.93,4.29) -- cycle    ;
\draw [shift={(92,290)}, rotate = 0] [fill={rgb, 255:red, 0; green, 0; blue, 0 }  ][line width=0.08]  [draw opacity=0] (8.93,-4.29) -- (0,0) -- (8.93,4.29) -- cycle    ;
\draw    (295,84) -- (295,89) -- (295,287) ;
\draw [shift={(295,290)}, rotate = 270] [fill={rgb, 255:red, 0; green, 0; blue, 0 }  ][line width=0.08]  [draw opacity=0] (10.72,-5.15) -- (0,0) -- (10.72,5.15) -- (7.12,0) -- cycle    ;
\draw    (459.33,125.67) -- (454.17,130.83) -- (297.12,287.88) ;
\draw [shift={(295,290)}, rotate = 315] [fill={rgb, 255:red, 0; green, 0; blue, 0 }  ][line width=0.08]  [draw opacity=0] (10.72,-5.15) -- (0,0) -- (10.72,5.15) -- (7.12,0) -- cycle    ;

\draw (284,60.67) node [anchor=north west][inner sep=0.75pt]   [align=left] {$\displaystyle \mathcal{T}_{A}$};
\draw (462,99.67) node [anchor=north west][inner sep=0.75pt]   [align=left] {$\displaystyle \mathcal{T}_{B}$};
\draw (74,306) node [anchor=north west][inner sep=0.75pt]   [align=left] {$\displaystyle m_{A}<0$};
\draw (526,305) node [anchor=north west][inner sep=0.75pt]   [align=left] {$\displaystyle m_{A}>0$};
\draw (271,304) node [anchor=north west][inner sep=0.75pt]   [align=left] {$\displaystyle m_{A}$ = 0};
\draw (306,177.33) node [anchor=north west][inner sep=0.75pt]   [align=left] {$\displaystyle g_{A,a}$};
\draw (127,225.67) node [anchor=north west][inner sep=0.75pt]   [align=left] {$\displaystyle \widehat{\mathcal{T}}_{A,-}$$\displaystyle \cong \widehat{\mathcal{T}}_{B,+}$};
\draw (406,227.67) node [anchor=north west][inner sep=0.75pt]   [align=left] {$\displaystyle \widehat{\mathcal{T}}_{A,+}$$\displaystyle \cong \widehat{\mathcal{T}}_{B,-}$};

\end{tikzpicture}

\caption{Our conjectured phase diagram, as a function of gauge couplings and the universal mass deformation, for an Abelian 3d $\CN=4$ gauge theory, $\CT_A$, with some arbitrary number of $U(1)$ gauge group factors and arbitrary number of integer charge hypermultiplets (we assume throughout that there are no decoupled $U(1)$ factors). Here the vertical direction represents flowing in the $\CN=4$ gauge coupling(s), $g_{A,a}$ ($a=1,\cdots,N_c$ runs over the individual $U(1)$ gauge group factors). The horizontal direction corresponds to the universal mass deformation, $m_A$. We argue that there is a single second-order phase transition, at $m_A=0$, which is also described by the IR of a dual UV gauge theory, $\widehat\CT_B$ (with $m_B=0$). We argue that the TQFTs on either side of $m_A=0$, $\widehat\CT_{A,\pm}$ are dual to those on either side of $m_B=0$, $\widehat\CT_{B,\mp}$, thereby providing evidence for the above phase diagram.}
\label{AbPhase}
\end{figure}
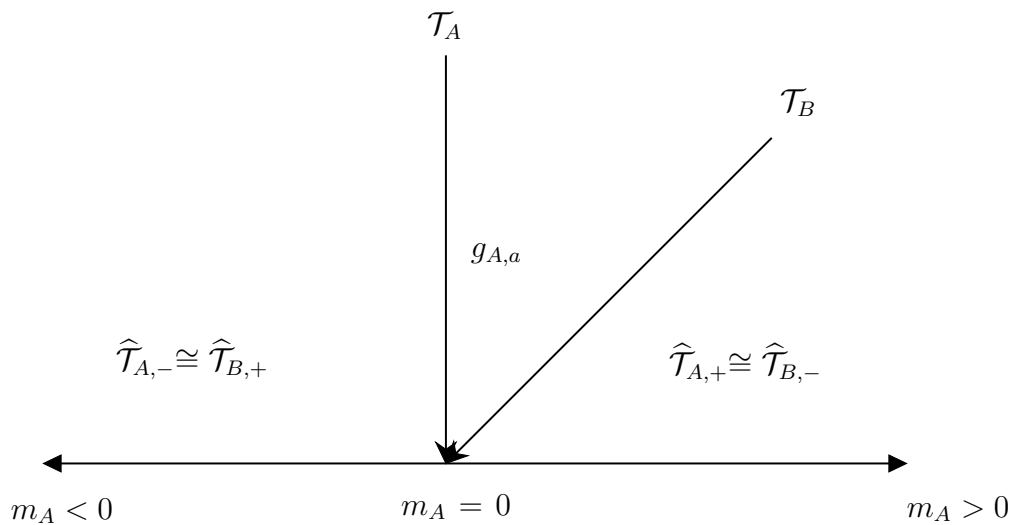

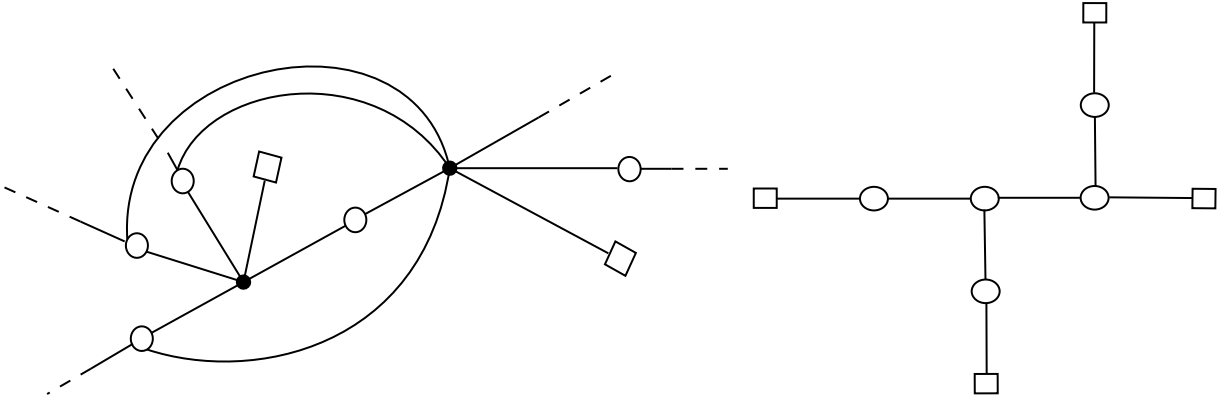
\begin{figure}[h!]
\centering

\tikzset{every picture/.style={line width=0.75pt}} 

\begin{tikzpicture}[x=0.75pt,y=0.75pt,yscale=-1,xscale=1]

\draw   (388.42,111.6) -- (400.01,111.6) -- (400.01,121.39) -- (388.42,121.39) -- cycle ;
\draw    (400.18,116.71) -- (441.98,116.71) ;
\draw   (441.98,116.71) .. controls (441.98,113.42) and (445.14,110.75) .. (449.03,110.75) .. controls (452.92,110.75) and (456.08,113.42) .. (456.08,116.71) .. controls (456.08,120) and (452.92,122.67) .. (449.03,122.67) .. controls (445.14,122.67) and (441.98,120) .. (441.98,116.71) -- cycle ;
\draw    (456.08,116.71) -- (497.88,116.71) ;
\draw   (497.88,116.71) .. controls (497.88,113.42) and (501.04,110.75) .. (504.93,110.75) .. controls (508.83,110.75) and (511.99,113.42) .. (511.99,116.71) .. controls (511.99,120) and (508.83,122.67) .. (504.93,122.67) .. controls (501.04,122.67) and (497.88,120) .. (497.88,116.71) -- cycle ;
\draw    (511.48,116.28) -- (553.28,116.28) ;
\draw   (553.28,116.28) .. controls (553.28,112.99) and (556.44,110.33) .. (560.34,110.33) .. controls (564.23,110.33) and (567.39,112.99) .. (567.39,116.28) .. controls (567.39,119.57) and (564.23,122.24) .. (560.34,122.24) .. controls (556.44,122.24) and (553.28,119.57) .. (553.28,116.28) -- cycle ;
\draw   (621.22,121.62) -- (609.63,121.52) -- (609.75,111.73) -- (621.33,111.83) -- cycle ;
\draw    (609.53,116.41) -- (567.73,116.05) ;
\draw    (504.73,122.22) -- (505.29,157.54) ;
\draw   (505.29,157.54) .. controls (509.18,157.49) and (512.38,160.12) .. (512.44,163.41) .. controls (512.49,166.7) and (509.37,169.41) .. (505.48,169.45) .. controls (501.59,169.49) and (498.39,166.86) .. (498.33,163.57) .. controls (498.28,160.28) and (501.4,157.58) .. (505.29,157.54) -- cycle ;
\draw   (499.89,215) -- (499.85,205.21) -- (511.43,205.18) -- (511.47,214.97) -- cycle ;
\draw    (505.89,205.05) -- (505.76,169.73) ;
\draw    (560.81,110.77) -- (560.47,75.46) ;
\draw   (560.47,75.46) .. controls (556.58,75.48) and (553.4,72.84) .. (553.36,69.55) .. controls (553.33,66.26) and (556.46,63.57) .. (560.36,63.54) .. controls (564.25,63.51) and (567.43,66.16) .. (567.46,69.45) .. controls (567.5,72.74) and (564.36,75.43) .. (560.47,75.46) -- cycle ;
\draw   (566.23,18.02) -- (566.21,27.8) -- (554.62,27.79) -- (554.64,18) -- cycle ;
\draw    (560.16,27.94) -- (560.08,63.26) ;
\draw    (84.69,184.46) -- (131.08,158.83) -- (182.58,130.37) ;
\draw    (82.14,143.47) -- (131.08,158.83) ;
\draw    (103.05,113.3) -- (131.08,158.83) ;
\draw    (141.8,107.6) -- (131.08,158.83) ;
\draw [shift={(131.08,158.83)}, rotate = 101.81] [color={rgb, 255:red, 0; green, 0; blue, 0 }  ][fill={rgb, 255:red, 0; green, 0; blue, 0 }  ][line width=0.75]      (0, 0) circle [x radius= 3.35, y radius= 3.35]   ;
\draw   (74.89,190.32) .. controls (73.44,187.29) and (74.46,183.53) .. (77.17,181.91) .. controls (79.88,180.3) and (83.24,181.43) .. (84.69,184.46) .. controls (86.14,187.48) and (85.12,191.24) .. (82.42,192.86) .. controls (79.71,194.48) and (76.34,193.34) .. (74.89,190.32) -- cycle ;
\draw   (74.56,145.77) .. controls (71.89,144.07) and (70.97,140.28) .. (72.49,137.31) .. controls (74.02,134.33) and (77.41,133.3) .. (80.08,135) .. controls (82.74,136.7) and (83.67,140.49) .. (82.14,143.47) .. controls (80.62,146.44) and (77.22,147.47) .. (74.56,145.77) -- cycle ;
\draw   (95.52,110.75) .. controls (94.07,107.73) and (95.09,103.97) .. (97.8,102.35) .. controls (100.51,100.73) and (103.88,101.87) .. (105.32,104.89) .. controls (106.77,107.92) and (105.75,111.68) .. (103.05,113.3) .. controls (100.34,114.91) and (96.97,113.78) .. (95.52,110.75) -- cycle ;
\draw   (138.96,92.93) -- (150.23,96.01) -- (147.47,108.59) -- (136.2,105.51) -- cycle ;
\draw   (182.58,130.37) .. controls (181.13,127.35) and (182.15,123.59) .. (184.86,121.97) .. controls (187.57,120.36) and (190.93,121.49) .. (192.38,124.52) .. controls (193.83,127.54) and (192.81,131.3) .. (190.11,132.92) .. controls (187.4,134.54) and (184.03,133.4) .. (182.58,130.37) -- cycle ;
\draw    (235.1,101.34) -- (192.38,124.52) ;
\draw [shift={(235.1,101.34)}, rotate = 151.52] [color={rgb, 255:red, 0; green, 0; blue, 0 }  ][fill={rgb, 255:red, 0; green, 0; blue, 0 }  ][line width=0.75]      (0, 0) circle [x radius= 3.35, y radius= 3.35]   ;
\draw    (279.96,75.72) -- (235.1,101.34) ;
\draw    (82.42,192.86) .. controls (135.95,210.51) and (221.33,192.43) .. (235.1,101.34) ;
\draw    (97.8,102.35) .. controls (110.69,58.64) and (197.37,43.84) .. (235.1,101.34) ;
\draw    (72.49,137.31) .. controls (66.34,44.98) and (216.23,14.81) .. (235.1,101.34) ;
\draw  [dash pattern={on 4.5pt off 4.5pt}]  (279.96,75.72) -- (320.2,52.5) ;
\draw    (235.1,101.34) -- (319.63,101.34) ;
\draw    (235.1,101.34) -- (315.1,144.32) ;
\draw   (318.64,138.3) -- (328.98,144.11) -- (323.7,155.67) -- (313.37,149.86) -- cycle ;
\draw   (325.58,95.67) .. controls (328.69,95.59) and (331.28,98.26) .. (331.38,101.65) .. controls (331.48,105.04) and (329.04,107.86) .. (325.94,107.94) .. controls (322.83,108.03) and (320.24,105.35) .. (320.14,101.97) .. controls (320.04,98.58) and (322.48,95.76) .. (325.58,95.67) -- cycle ;
\draw    (331.38,101.65) -- (346.96,101.65) ;
\draw  [dash pattern={on 4.5pt off 4.5pt}]  (346.96,101.65) -- (375.33,101.65) ;
\draw    (75.32,189.98) -- (54.12,202.53) ;
\draw  [dash pattern={on 4.5pt off 4.5pt}]  (54.12,202.53) -- (32.04,215.33) ;
\draw    (71.15,138.31) -- (48.86,128.29) ;
\draw  [dash pattern={on 4.5pt off 4.5pt}]  (48.86,128.29) -- (9.5,110.58) ;
\draw    (92.89,93.53) -- (97.8,102.35) ;
\draw  [dash pattern={on 4.5pt off 4.5pt}]  (65.39,51.15) -- (89.73,88.93) ;

\end{tikzpicture}

\caption{A generic graph corresponding to one of the Abelian gauge theories we study on the left and a quiver gauge theory that we also study on the right (we need only analyze connected graphs). In the generic theory, circles represent $U(1)$ gauge group factors, while black vertices denote hypermultiplets charged under up to $N_c$ different $U(1)$ gauge groups. The lines connecting black dots and circles represent charges (each corresponding to an entry in a matrix, $q_{ai}$, with $i=1,\cdots,N_f$ running over hypermultiplets, and $a=1,\cdots, N_c$ running over gauge groups). The white boxes represent flavor symmetries (although the flavor charges are subject to a constraint that the gauge / flavor matrix, $\mathbf{q}\in GL(N_f,\mathbb{Z})$). On the right we have the special case of a quiver theory. In this case, all matter is fundamental or bifundamental and black dots become superfluous. Instead, each $U(1)$ factor corresponds to a circle and edges correspond to (bi)fundamental hypermultiplets. Squares again correspond to flavor nodes (subject to the same $\mathbf{q}\in GL(N_f,\mathbb{Z})$ constraint as in the general case).}
\label{AbGraph}
\end{figure}

The purpose of this paper is to argue the following:
\begin{enumerate}
\item[{\bf A1:}] The answer to the first question turns out to be familiar from the physics of the fractional quantum Hall effect (FQHE). The answer can also be arrived at by thinking in terms of defects \cite{Barkeshli:2014cna}. The main point is the following: due to the continuous nature of the flavor symmetries in question, they cannot permute the lines of the IR TQFT. However, we can insert lines of the IR TQFT at junctions of continuous symmetry defects (as in Fig. \ref{FracFig}). This maneuver leads to a notion of symmetry fractionalization that allows us to reproduce the UV 't Hooft anomalies in the IR TQFT and, in many cases, assign a Hall conductance to the IR phase (e.g., see \cite{Cheng:2022nds,Cheng:2022nji}).

For example, in the case of 3d $\CN=4$ SQED with $N_f$ flavors of charge $1$, we have a mixed 't Hooft anomaly between the $U(1)_T$ symmetry (with Noether current $\star F$) and the $PSU(N_f)$ flavor symmetry.\footnote{Therefore, our TQFT anomaly matching computation  shows that the gapped phase carries some information about both the Coulomb and Higgs branches of the UV theory (since these moduli spaces have an action of $U(1)_T$ and $PSU(N_f)$ on them respectively).} In the IR we show that, up to the action of invertible (accidental) discrete symmetries, this 't Hooft anomaly can be reproduced via a unique fractionalization class in the TQFT (all different fractionalization classes related by these discrete symmetries lead to the same Hall conductance).\footnote{The mixed 't Hooft anomaly requires $N_f>1$ in order to have a non-trivial $PSU(N_f)$ factor. We will comment on the $N_f=1$ case separately.}

Extending these anomaly matching ideas, we also comment on some properties of TQFTs arising from universal deformations of more abstract and general local unitary 3d $\CN=4$ SCFTs.
\item[{\bf A2:}] In the non-conformal UV regime of a 3d $\CN=4$ gauge theory that flows to an SCFT, we would like to think of the ancestor of the deformation in \eqref{univDef} as a deformation that preserves 3d $\CN=4$ SUSY but gives flavor and $\frak{so}(4)_R$-preserving (see footnote \ref{RsymmPres}) masses to all the degrees of freedom. Already in the simple (though admittedly special) case of a free vector multiplet, this desire leads to a puzzle. In this theory, the only way to give mass to the gauge field is a Chern-Simons (CS) term. However, the lore is that such terms are not compatible with $\CN=4$ SUSY. Indeed, while $\CN=4$ BF terms are well known (by coupling vector multiplets to twisted vector multiplets) \cite{Kapustin:1999ha}, the status of SUSY completions of standard CS terms is somewhat less known (although see the example in Appendix E.1 of \cite{Lin:2005nh}).

In the case of pure Abelian gauge theories (without matter), we explicitly show such terms are indeed allowed, thereby extending the analysis of Gaiotto and Witten for the case of Chern-Simons-matter theories \cite{Gaiotto:2008sd} (see also \cite{Assel:2022row}) with vanishing kinetic terms. As a particularly simple example, we argue that, in the case of $\CN=4$ super-Maxwell theory, such a CS term deforms the supersymmetry algebra as follows\footnote{We also comment on generalizations of this discussion to arbitrary Abelian theories in the non-conformal UV.}
\begin{equation}\label{MaxwellSUSY}
\left\{Q_{\alpha}^{a\dot a},Q_{\beta}^{b\dot b}\right\}=\varepsilon^{ab}\varepsilon^{\dot a\dot b}P_{(\alpha\beta)}-{g^2k\over8\pi}\varepsilon_{\alpha\beta}\left(\varepsilon^{ab}R_C^{\dot a\dot b}-\varepsilon^{\dot a\dot b}R_H^{ab}\right)~,
\end{equation}
where $R_{C}$ ($R_H$) are the $\mathfrak{su}(2)_C$ ($\mathfrak{su}(2)_H$) $R$-symmetry currents, $g$ is the gauge coupling, and $k$ is the CS level. This theory flows to a pure $U(1)_k$ CS theory in the deep IR.\footnote{In general, even when $k$ is even, there will be a transparent fermion present; we will comment on this fact in more detail in the body of the paper.} The algebra \eqref{MaxwellSUSY} is essentially the same one that arises under the deformation in \eqref{univDef} \cite{Cordova:2016xhm}
\begin{equation}\label{defAlg}
\left\{Q_{\alpha}^{a\dot a},Q_{\beta}^{b\dot b}\right\}=\varepsilon^{ab}\varepsilon^{\dot a\dot b}P_{(\alpha\beta)}+m\varepsilon_{\alpha\beta}\left(\varepsilon^{ab}R_C^{\dot a\dot b}-\varepsilon^{\dot a\dot b}R_H^{ab}\right)~,
\end{equation}
where, in \eqref{MaxwellSUSY}, $m=-{g^2k/8\pi}$.

In what follows, we are interested in Abelian 3d $\CN=4$ gauge theories with matter and the SCFTs that arise from them via flows in the gauge coupling(s). As we will describe in more detail in our answer to {\bf Q3}, our prescription for the UV ancestor of the universal deformation in these theories involves a mix of tree-level (in the matter) and one-loop (in the gauge multiplets) physics.

\item[{\bf A3:}]  We will argue that our theories of interest have simple phase diagrams of the type depicted in Fig. \ref{AbPhase}. In the UV, we have two mirror pairs of gauge theories ($\CT_A$ and $\CT_B$) that, in the absence of mass deformations, flow to a common IR SCFT. Then, if we turn on the relevant deformation in \eqref{univDef}, we get two dual spin TQFTs (in a sense that includes and generalizes standard level-rank dualities)
\begin{eqnarray}\label{dualityAB}
{\rm CS}(\widehat K_A)\ &\leftrightarrow&\ {\rm CS}(\widehat K_B)~.
\end{eqnarray}
Here, $\widehat K_{A}$ ($\widehat K_B)$ is the $K$-matrix corresponding to the IR CS theory with action\footnote{Note that the factor of $i$ in front of the CS terms arises because we are working in Euclidean signature.}
\begin{equation}\label{Kmat}
S_{{\rm CS}(K)}={i\over4\pi}\int d^3x\vec a^T\cdot K\cdot d \vec a~.
\end{equation}
Up to stacking with an SPT we will describe in the main text, the entries in the $K$ matrix are generated at 1-loop by integrating out matter fields. In particular, $K$ captures the graph that specifies $\CT$ as an Abelian gauge theory (see Fig. \ref{AbGraph}). The $K$ matrices are sensitive to the sign of the universal masses, $m_{A,B}$, that we turn on. In particular, the signs of the $K$ matrices change when the signs of the universal masses change. Moreover, as we will discuss in the main text, $m_A\sim-m_B$ have opposite signs.

In order to make the notion of 1-loop CS terms precise, we must in fact give masses to the fields in the non-conformal UV gauge theory (in the deep IR, where the gauge coupling is strong, there is no notion of a loop expansion). In other words, to carry out the computation leading to \eqref{Kmat}, we take the limit $|m|\gg g_{I}^2$ for all gauge couplings, $g_{I}$. In this limit, mirror symmetry does not operate (it is an IR duality), and, as described in {\bf A2}, we must make sense of a non-conformal version of \eqref{univDef}.

For a general Abelian gauge theory, our proposal is as follows. We set all UV CS terms to zero and then integrate out the matter fields in $\CT_A$ with mass $|m_A|\gg g_{A,a}^2$ (the $g_{A,a}$ are the gauge couplings of $\CT_A$) at tree level 
\begin{equation}\label{mattDef}
\delta S\sim\int d^3x m_A\ \varepsilon^{\alpha\beta}\varepsilon_{ab}\varepsilon_{\dot a\dot b}Q^{a\dot a}_{\alpha}Q^{b\dot b}_{\beta}J_{\rm matter}+\CO(m_A^2)~,
\end{equation}
where $J_{\rm matter}$ is the primary of the stress tensor multiplet of the (weakly coupled) UV matter fields. This procedure breaks $\CN=4$ supersymmetry, and upon integrating out the matter fields, we generate an effective theory of fields from Abelian $\CN=4$ vector multiplets with one-loop CS terms (in addition to kinetic terms for the vector multiplets)
\begin{equation}\label{effCS}
\delta S\sim {i\over4\pi}\int d^3x\vec a^T\cdot K_A\cdot d\vec a+\cdots~,
\end{equation}
where the ellipses includes a completion we describe in the main text. We expect this completion preserves a generalization of the deformed $\CN=4$ algebra in \eqref{MaxwellSUSY} as we lower $|m|g_{A_a}^{-2}$. Thus, as we lower $|m|g_{A_a}^{-2}$, the deformed SUSY algebra is emergent, and we can think of the combined tree-level \eqref{mattDef} and one-loop \eqref{effCS} deformations as comprising the UV analog of the universal deformation \eqref{univDef}.\footnote{Another proposal would be to take the non-conformal UV stress tensor multiplet and deform the UV theory by it. We do not expect this approach to give rise to a phase diagram qualitatively different from ours (since we expect this deformation to also give mass to all the fields in the theory). Note that, in the somewhat simpler case of Chern-Simons-matter theories, \cite{Hosomichi:2008jd} considers a deformation essentially equivalent to ours (in their case, the gauge degrees of freedom are already, in some sense, topological and so do not \lq\lq contribute" to the EM tensor; therefore, we can also think of their deformation as being equivalent to deforming by the full UV stress tensor). See also \cite{Abel:2011wv,Buican:2012ec,Cordova:2018acb} for other contexts where deformations by stress tensor multiplets played an imporant role. \label{Tproposal}} In particular, we associate the algebra in \eqref{MaxwellSUSY} with the deformed SCFT algebra of \eqref{univDef} (in other words, we believe there is no phase transition between the IR SCFT deformed as in \eqref{univDef} and the phase governed by \eqref{MaxwellSUSY}).

We then flow to the IR and find the dual theories in \eqref{dualityAB} when deforming $\CT_B$ by $|m_B|\gg g^2_{B,i}$ (with opposite sign $m_B$). The fact that these dualities hold is strong evidence that there is no further phase transition as we lower $|m_A|$ relative to $g^2_{A_i}$ beyond the second-order one at zero universal mass.\footnote{We prove the dualities in \eqref{dualityAB} for the case of general Abelian gauge theories with no one-form symmetry. We then argue how this proof generalizes to the case with one-form symmetry and explicitly check our argument in infinitely many examples.}

\end{enumerate}

\bigskip
The plan of this paper is as follows. In the next section, we review basic properties of 3d $\CN=4$ SUSY with a special emphasis on mirror symmetry in the context of Abelian gauge theories. We also explain more precisely our proposal for a UV ancestor of the universal mass deformation. Then, in Sec. \ref{no1form} we describe the phase diagram for the universal mass deformation of any $\CN=4$ Abelian gauge theory with compact gauge group and no 1-form symmetry. Along the way, we exhibit duality of all corresponding mirror IR TQFT pairs. Moreover, in Sec. \ref{AnomMatchSec} we explain our answer to {\bf Q1} and highlight how anomaly matching for continuous symmetries works in 3d $\CN=4$ SQED and its dual. In Sec. \ref{1formResults} we generalize our results to $\CN=4$ Abelian gauge theory with 1-form symmetry. Together with Sec. \ref{no1form}, this material constitutes our answers to {\bf Q2} and {\bf Q3}. Then, in Sec. \ref{general}, we revisit and generalize aspects of our anomaly matching discussion from Sec. \ref{AnomMatchSec} and explain how any local 3d $\CN=4$ SCFT with certain anomalies for a $U(1)\times PSU(N)$ symmetry can be deformed to a theory that includes an IR TQFT with a factor described by an Abelian Chern-Simons theory. We conclude with a discussion of open questions.

\newsec{3d $\CN=4$ (deformed) supersymmetry, mirror symmetry, and the universal mass}\label{generalDef}
We are mainly interested in theories with 3d $\CN=4$ SUSY, although there are close non-SUSY (and $\CN<4$ SUSY) analogs for many of our statements. In this section, we very briefly review some relevant facts about 3d $\CN=4$ QFTs. 

The main focus of our work is on understanding what happens when we turn the universal deformation in \eqref{univDef} on in a local unitary 3d $\CN=4$ SCFT. To understand this deformation at a somewhat more technical level, we note that $J$ in \eqref{univDef} is the scaling dimension-one primary for the multiplet that contains the EM tensor of a local 3d $\CN=4$ SCFT (see \cite{Cordova:2016emh,Cordova:2016xhm} for a full list of 3d $\CN=4$ multiplets and deformations). At linear order in $m$, this universal deformation involves the action of two Poincar\'e supercharges, $Q_{\alpha}^{a\dot a}$ (rendering the linear deformation scaling dimension two), which transform as follows under the symmetries of such a theory
\begin{equation}\label{Qsymm}
Q_{\alpha}^{a\dot a}\in {\bf(2,2,2)}\ {\rm of}\ \mathfrak{su}(2)_C\oplus \mathfrak{su}(2)_H\oplus \mathfrak{su}(2)~.
\end{equation}
In this expression, the first two groups on the RHS form the $\mathfrak{so}(4)_R$ $R$ symmetry. Typically, a 3d $\CN=4$ SCFT also has additional continuous bosonic flavor symmetries that commute with the supercharges (but the above symmetries are universal). The universal deformation preserves all these flavor symmetries and also the bosonic symmetries appearing in \eqref{Qsymm}.

Intuitively, if we turn on a relevant deformation proportional to an operator in the EM tensor multiplet, we expect all degrees of freedom in the theory to obtain a mass (since, in a Lagrangian theory, all fields \lq\lq contribute" to the EM tensor\footnote{As a simple illustration, we explicitly demonstrate \eqref{defAlgM} in the case of the SCFT corresponding to the free massless hypermultiplet (see Appendix \ref{Hyperexample}). }). At the level of the supersymmetry algebra, turning on the deformation in \eqref{univDef} (along with various higher-order in $m$ deformations needed to preserve SUSY) results in the following deformation of the $\CN=4$ algebra \cite{Cordova:2016xhm}
\begin{equation}\label{defAlgM}
\left\{Q_{\alpha}^{a\dot a},Q_{\beta}^{b\dot b}\right\}=\varepsilon^{ab}\varepsilon^{\dot a\dot b}P_{(\alpha\beta)}\ \longrightarrow\ \left\{Q_{\alpha}^{a\dot a},Q_{\beta}^{b\dot b}\right\}=\varepsilon^{ab}\varepsilon^{\dot a\dot b}P_{(\alpha\beta)}+m\varepsilon_{\alpha\beta}\left(\varepsilon^{ab}R_C^{\dot a\dot b}-\varepsilon^{\dot a\dot b}R_H^{ab}\right)~.
\end{equation}
Then, the abstract arguments in \cite{Cordova:2016xhm,Lin:2005nh} show that an IR theory with such a deformed SUSY algebra is necessarily gapped. We  expect this gapped theory to be typically described by a non-trivial TQFT that is sensitive to the sign of the universal mass.

In practice, understanding the universal mass deformation directly in an SCFT is difficult (except for the case of collections of free hypermultiplets and other closely related theories), because most such theories are usually strongly coupled. A somewhat more practical approach is to understand how to describe an analog of the universal mass deformation in a UV $\CN=4$ gauge theory that flows to an SCFT in the IR. The UV matter degrees of freedom of such a theory sit inside hypermultiplets
\begin{equation}\label{hyper}
\rho^{a,i}\in {\bf(1,2,1)}~,\ \psi_{\alpha}^{\dot a, i}\in {\bf(2,1,2)}~,
\end{equation}
where $\dot a$ and $a$ are $\mathfrak{su}(2)_C$ and $\mathfrak{su}(2)_H$ $R$ symmetry spinor indices respectively, $\alpha$ is an $\mathfrak{su}(2)$ spinor index, and $i$ is an index for a representation, $\pi_G$, of the UV gauge group under which the matter fields in question transform. The gauge bosons sit inside vector multiplets
\begin{equation}\label{vector}
\Phi^{(\dot a\dot b),A}\in {\bf(3,1,1)}~,\ \lambda_{\alpha}^{a\dot a, A}\in {\bf(2,2,2)}~,\ F^A_{(\alpha\beta)}\in{\bf(1,1,3)}~, \ D^{(ab),A}\in{\bf(1,3,1)}~,
\end{equation}
where $a$, $\dot a$, $b$, and $\dot b$ are $R$ symmetry indices, $\alpha$ and $\beta$ are $\mathfrak{su}(2)$ indices, and $A$ is an adjoint index of the gauge group. The vector multiplets then couple to the matter fields via the supersymmetric completions of the linear couplings of gauge fields to matter, $\delta\CL\supset A^A_{\mu}J^{\mu}_A+\cdots$ (where we must also include the usual seagull terms to maintain gauge invariance).

However, there are various issues that arise in the UV gauge theory regime: the QFTs are not conformal (and so it is unclear if SUSY can even be preserved under an ancestor of the universal deformation), and the UV gauge theory description is not unique (i.e., many distinct UV gauge theories may flow to the same IR SCFT). We will deal with the first issue in the next section. The second issue presents an opportunity, because it gives us consistency conditions for the TQFTs arising from the universal deformation of the IR SCFT,  and it also helps explain how to fit the deformed IR SCFT into a broader phase diagram.

Mirror symmetry \cite{Intriligator:1996ex} is a canonical source of non-uniqueness in the UV Lagrangian description. Very roughly, mirror symmetry is an IR equivalence between two distinct UV gauge theories that flow to the same SCFT at long distances. This duality exchanges
\begin{enumerate}
\item $\mathfrak{su}(2)_C$ and $\mathfrak{su}(2)_H$
\item Higgs branches of vacua and Coulomb branches of vacua.\footnote{This exchange follows from the fact that these moduli spaces of vacua are always parameterized by expectation values of $\mathfrak{su}(2)_H$-charged and $\mathfrak{su}(2)_C$-neutral matter operators and $\mathfrak{su}(2)_H$-neutral and $\mathfrak{su}(2)_C$-charged vector multiplet / (dressed) monopole operators respectively (this statement holds in all known theories and is conjectured to hold in all local unitary $\CN=4$ QFTs). More precisely, the Higgs and Coulomb branches, $X_H$ and $X_C$ respectively, arise as $X_{H,C}:={\rm Spec}(\CC_{H,C})$, where $\CC_{H,C}$ are the corresponding chiral rings of half-BPS operators having the charge assignments described in this footnote. \label{SwapPrimary}}
\item Flavor symmetries (and corresponding mass parameters) that act on the Higgs branch and flavor symmetries (and corresponding mass parameters) that act on the Coulomb branch.\footnote{In many theories, this fact can be related to the discussion in footnote \ref{SwapPrimary} via the swapping of scaling dimension one primaries transforming as ${\bf(3,1)}$ irreps under $\mathfrak{su}(2)_C\oplus \mathfrak{su}(2)_H$ with those transforming as ${\bf(1,3)}$ (since the resulting multiplets host Noether currents for the corresponding flavor symmetries). A particularly simple example of this phenomenon is the supersymmetric completion of the exchange of the \lq\lq topological" $U(1)$ symmetry (with $\star F$ Noether current) and an Abelian symmetry acting on mirror matter fields in Lagrangian theories.}
\end{enumerate}
For us, a particularly important (and almost tautological) additional entry is the following:
\begin{enumerate}
\item[4.] The stress tensor multiplet must be mapped to itself. This statement is compatible with the fact that the primary, $J$, is neutral under the $R$ symmetry. The corresponding universal mass deformation is then mapped to itself up to a sign (we will return to discuss this sign below). 
\end{enumerate}
\paragraph{}

Much of our discussion in subsequent sections will focus on the case of 3d $\CN=4$ Abelian gauge theories. In this context, we have a $U(1)^{N_c}$ gauge group with $N_f$ hypermultiplets. We can encode the corresponding charges of the hypermultiplets in the $N_c\times N_f$ integer matrix, $q_{ai}$, with $a=1,\cdots,N_c$ labelling the gauge group and $i=1,\cdots N_f$ labelling the hypermultiplets. 
 We take the corresponding graph (see Fig. \ref{AbGraph}) to be connected (since we can, without loss of generality, focus on \lq\lq irreducible" theories) and so $N_c\le N_f$. However, we are motivated to complete this matrix into an $N_f\times N_f$ integer matrix\footnote{This matrix is natural because we are effectively constructing a vector space involving $N_f$ independent charge vectors (the $\CN=4$ gauge couplings do not furnish constraints).}
\begin{equation}\label{Aside}
    \mathbf{q} = \begin{pmatrix}
        q \\
        \hat q
    \end{pmatrix}~,
\end{equation}
where $\hat q_{\alpha i}$ is an $(N_f-N_c)\times N_f$ matrix of $U(1)$ flavor charges. By construction, $\mathbf{q}$ is invertible and therefore has non-zero determinant. An additional physical input that we will demand in the next section (but then relax in the subsequent one) is that ${\rm det}(\mathbf{q})=\pm1$ (i.e., $\mathbf{q}\in GL(N_f,\mathbb{Z})$). This condition rules out the existence of 1-form symmetry.\footnote{We may understand this statement as follows. We have one-form symmetry if at least a $U(1)$ subgroup of the gauge group has hypermultiplets with charges $n_i q'$ with $n_i,2\le q'\in\mathbb{Z}$ (in this case, there is a non-trivial $\mathbb{Z}_{q'}<U(1)$ one-form symmetry remaining). Then, $|{\rm det}(\mathbf{q})|\ge q'>1$. \label{1-form}}

In Abelian theories, the action of mirror symmetry on the corresponding Lagrangians takes a relatively simple form \cite{deBoer:1996ck,Bullimore:2015lsa}. Indeed, one finds a $U(1)^{N_f-N_c}$ gauge theory with
\begin{equation}\label{Bside}
\mathbf{q}\to\tilde{\mathbf{q}}=\begin{pmatrix}
        \tilde{\hat q} \\
        \tilde q
    \end{pmatrix}~,\ \ \ \tilde{\mathbf{q}}=\mathbf{q}^{-1,T}~,
\end{equation}
where $\tilde q$ is the $(N_f-N_c)\times N_f$ matrix of mirror gauge charges (here $\tilde{\hat{q}}$ is an $N_c\times N_f$ matrix of flavor charges). As an aside, we therefore also see that when ${\rm det}({\mathbf{q}})=\pm1$, not only do we have the absence of 1-form symmetry, but we also find that the mirror gauge charges are integers. 

Both the $U(1)^{N_c}$ gauge theory corresponding to \eqref{Aside} and the $U(1)^{N_f-N_c}$ gauge theory corresponding to \eqref{Bside} are believed to flow to the same IR SCFT. At this fixed point, we would like to turn on the universal deformation \eqref{univDef} in both duality frames and flow to dual TQFTs. In addition to being of independent physical interest, such a flow may provide further evidence that the original duality between non-topological QFTs is correct.

However, the IR theories in question are typically strongly coupled, and so it is difficult to study their behavior under the universal deformation. Instead, we will turn on a UV ancestor of the universal deformation. This deformation should preserve the $R$ symmetry and flavor symmetry while giving mass to all the fields in the problem. This discussion also presumes that the UV theory has the full $\CN=4$ $R$ symmetry (e.g., this means that we do not have any FI parameters or flavor mass terms turned on).

For free hypermultiplets, such a deformation is simple enough to arrange. We have
\be\label{hyperUniv}
S=\int d^3 x \left(\p^\mu \tilde   \rho^a \p_\mu  \rho_a -i \tilde \psi^{\dot a}\gamma^\mu  \p_\mu \psi_{\dot a}+{m^2} \tilde   \rho^a  \rho_a+im\tilde\psi^{\dot a}\psi_{\dot a}\right)~.
\ee
It is straightforward to check (see Appendix \ref{Hyperexample}) that the deformed algebra in this case matches that in \eqref{defAlg}. 

For the free vector multiplet, the situation is somewhat more subtle. In order to generate a mass for the gauge boson, we must turn on a CS term. We claim that, even for a single $U(1)$ gauge field, we can complete this deformation into an $\CN=4$-invariant Lagrangian as follows
\begin{equation}\label{vectUniv}
\begin{aligned}
S= \int d^3x\Big(\frac{1}{g^2} \Big(&F^{\mu\nu}F_{\mu\nu} - D^\mu \Phi^{\da\db} D_\mu\Phi_{\da\db} + i\lambda^{\alpha a \db} (\gamma^\mu D_\mu \lambda)_{\alpha a \db}- D^{ab}D_{ab}\Big)\\
& + {ik\over4\pi} \varepsilon_{\mu\nu\rho} F^{\mu\nu}A^\rho - {ik\over4\pi} \lambda^{\beta a \db}\lambda_{\beta a \db}-{k^2g^2\over16\pi^2} \Phi^{\da\db}\Phi_{\da\db}\Big)~.\\
\end{aligned}
\end{equation}
As we show in Appendix \ref{MCSexample}, \eqref{vectUniv} leads to a theory with a deformed $\CN=4$ algebra with universal mass $m=-{kg^2\over8\pi}$ as in \eqref{MaxwellSUSY}. In the next subsection, we study how to define an ancestor of the universal mass deformation in gauge theories with matter.

\subsection{The universal mass deformation in Abelian gauge theories}
Given the above discussion, we would like to understand how to make sense of the universal deformation in the UV of more general Abelian gauge theories with matter. The ancestor of the universal deformation should give all matter fields the same mass, $m$. Therefore, prior to gauging, we should have
\be\label{hyperUnivGen}
S_{\rm matter}=\sum_i\int d^3 x \left(\p^\mu \tilde   \rho^{ai} \p_\mu  \rho_{ai} -i \tilde \psi^{\dot ai}\gamma^\mu  \p_\mu \psi_{\dot ai}+{m_A^2} \tilde   \rho^{ai}  \rho_{ai}+im_A\tilde\psi^{\dot ai}\psi_{\dot ai}\right)~,
\ee
where $i$ runs over the matter hypermultiplets. Similarly, in the gauge sector we could try to arrange for a common universal mass. For example, we could consider
\begin{equation}\label{vectUnivGen}
\begin{aligned}
S_{\rm gauge} =\sum_{c}\int d^3x\Big[ \frac{1}{g_{c}^2} \Big(&F_{c}^{\mu\nu}F_{c\mu\nu} - D^\mu \Phi_{c}^{\da\db} D_\mu\Phi_{c\da\db} + i\lambda_{c}^{\alpha a \db} (\gamma^\mu D_\mu \lambda_{c})_{\alpha a \db}- D_{c}^{ab}D_{c ab}\Big)\\
& + {ik_{c}\over4\pi} \varepsilon_{\mu\nu\rho} F_{c}^{\mu\nu}A_{c}^\rho - {ik\over4\pi} \lambda_{c}^{\beta a \db}\lambda_{c\beta a \db} - {k^2g^2\over16\pi^2} \Phi_{c}^{\da\db}\Phi_{c\da\db}\Big]~,\\
\end{aligned}
\end{equation}
where $c$ runs over the different gauge sectors (so each $k_{c}g_{c}^2=kg^2$ is independent of $c$). 

Clearly, the gauge and matter sectors each preserve a generally different (diagonal) SUSY algebra of the form \eqref{defAlg} (one involves $m=m_A$, and the other involves $m=-{kg^2\over8\pi}$). For these algebras to be compatible, we should set $m_A=-{kg^2\over8\pi}$
Then, by coupling the gauge multiplets in \eqref{vectUnivGen} to the matter fields in \eqref{hyperUnivGen}, we get a gauge theory that preserves the deformed algebra (see Appendix \ref{interactingApp} for further details)
\begin{equation}
\left\{Q_{\alpha}^{a\dot a},Q_{\beta}^{b\dot b}\right\}=\varepsilon^{ab}\varepsilon^{\dot a\dot b}P_{(\alpha\beta)}\ \longrightarrow\ \left\{Q_{\alpha}^{a\dot a},Q_{\beta}^{b\dot b}\right\}=\varepsilon^{ab}\varepsilon^{\dot a\dot b}P_{(\alpha\beta)}+m_A\varepsilon_{\alpha\beta}\left(\varepsilon^{ab}R_C^{\dot a\dot b}-\varepsilon^{\dot a\dot b}R_H^{ab}\right)~.
\end{equation}

However, the above theory is peculiar. First, by setting $m_A=-{kg^2\over8\pi}$, we see that, up to a discrete coupling (the CS level), there is a single continuous deformation parameter in the UV. On the other hand, we expect the UV ancestor of the IR universal deformation to furnish a second continuous deformation parameter (in addition to the gauge coupling) since it corresponds to a distinct continuous deformation of the IR SCFT. Second, as commented in \cite{Cordova:2016xhm}, without invoking some non-analytic properties of the $S$ matrix (perhaps as in \cite{Gabai:2022snc,Jain:2014nza}), the above theory looks somewhat puzzling from the point of view of \cite{Haag:1974qh}.

We can avoid both of the above complications simply by relaxing the requirement that the UV ancestor of the universal deformation preserves SUSY in the UV and instead allow it (and the corresponding deformed algebra) to be emergent in the IR (see Footnote \ref{Tproposal} for a discussion of other qualitatively similar possibilities). Which value of the UV CS parameter should we choose? Since the universal deformation of the IR SCFT is a single continuous deformation (with no associated discrete parameters), the most natural map to the UV is for us to take $k_{c}=0$ and take the CS terms to be those induced at one loop by integrating out the massive matter.

In other words, we start from the UV degrees of freedom as in \eqref{hyperUnivGen} and take \eqref{vectUnivGen} with $k_{c}=0$. Coupling the gauge fields to the matter fields, we go to the regime $|m_A|\gg g_{c}^2$, and integrate out the matter at one loop to obtain the symmetric $K$ matrix of CS couplings
\begin{equation}\label{A1loop}
(K_A)_{ab}={\rm sign}(m_A)\sum_{i=1}^{N_f} q_{ai}q_{bi}\ \Rightarrow\ K_A={\rm sign}(m_A)qq^T~.
\end{equation}
More generally, we should consider the extended symmetric $\widehat N_A\times\widehat N_A$ $K$ matrix
\begin{equation}\label{A1loopExt}
\widehat{K}_A=K_A\oplus \Pi_A~,
\end{equation}
where $\Pi_A$ is a diagonal matrix with entries $\pm1$. Here $\Pi_A$ corresponds to stacking with a $U(1)_1^{n_A}\boxtimes U(1)_{-1}^{\ell_a}$ SPT. 

In the deep IR, we find the following TQFT, $\widehat\CT_A$\footnote{Note that in Fig. \ref{AbPhase} we included two TQFTs, $\widehat\CT_{A,\pm}$, in the phase diagrams. In the context of our present discussion, this refinement corresponds to explicitly keeping track of the two possible signs for $m_A$ in \eqref{A1loop} (which we emphasize correspond to generically {\it different} TQFTs, although they are dual in certain cases). However, for simplicity, we will drop the \lq\lq$\pm$" subscript in what follows and keep track of the sign of $m_A$ implicitly.}
\begin{equation}\label{TQFTA}
S_{IR} = {i\over4\pi}\int d^3x\vec a^T \widehat K_A d\vec{a}~.
\end{equation}
If $K_A$ has even diagonal entries, then the extension $K_A\to\widehat K_A$ corresponds to tensoring in a transparent fermion (an ${\rm SVec}$ factor in the category theory language) that turns the non-spin TQFT into a spin TQFT.\footnote{Since our starting points are supersymmetric, we expect the IR TQFTs we generate to be spin TQFTs.} This extension of the $K$ matrix also shifts the central charge of the TQFT (which will be important in our discussion of TQFT dualities). If the theory is already a spin TQFT, taking $K_A\to\widehat K_A$ does not add a transparent fermion, but it does still shift the central charge.\footnote{This statement holds because each time we tensor in a $U(1)_{\pm1}$ factor to a spin TQFT, we also condense a diagonal $\CA=(1,1)+(f,f)$ algebra (where $(f,f)$ is a product of transparent fermions from the original theory and the added $U(1)_{\pm1}$). \label{CondenseTensor}} Note that, in writing \eqref{TQFTA}, we have neglected various couplings involving background fields (including the gravitational CS term). We will return to some of these terms when we consider anomaly matching in the next section.

To better understand some of these statements, let us consider the line operators of the spin TQFT, $\widehat\CT_A$ (see \cite{Delmastro:2019vnj} for a recent review)
\begin{equation}\label{lineOps}
\ell_{\vec\alpha}(\CC):=\exp\left(i\vec\alpha^T\cdot\int_{\CC}\vec a\right)~,
\end{equation}
where the charge vectors live in equivalences classes, $[\vec\alpha]\in\mathbb{Z}^{\widehat N_A}/\sim$. More precisely, we work in an $\widehat N_A$-dimensional integer lattice modulo the equivalence relation
\begin{equation}\label{LatticeID}
[\vec\alpha]=[\vec\beta] \ \ \ \Leftrightarrow\ \ \ \vec\alpha=\vec\beta+\widehat K_A\vec\gamma~.
\end{equation}
Here $\vec\gamma$ is any $\widehat N_A$-dimensional integer vector satisfying
\begin{equation}\label{GammaConstraint}
\sum_i(\widehat K_A)_{ii}\gamma_i\in 2\mathbb{Z}~.
\end{equation}
We say a vector satisfying \eqref{GammaConstraint} is \lq\lq $\widehat K_A$ even." This condition ensures that we obtain $2|\det\widehat K_A|=2|\det K_A|$ lines, which is the number we expect for a spin TQFT. In particular, if the TQFT with $K$ matrix $K_A$ is already a spin TQFT, \eqref{GammaConstraint} ensures that the extension $K_A\to\widehat K_A$ does not give rise to any additional lines (see also footnote \ref{CondenseTensor}). On the other hand, if $K_A$ is a non-spin TQFT (which is equivalent to $K_A$ having all even diagonal entries), then \eqref{GammaConstraint} ensures that we add in an additional transparent fermion and all tensor products of that fermion with lines in the theory defined by $K_A$. In other words, we extend our non-spin TQFT to a spin TQFT and double the number of lines.

From the above discussion we also conclude that the OPE of lines is Abelian
\begin{equation}\label{FusionRules}
\ell_{\vec\alpha}\times\ell_{\vec\beta}=\ell_{\vec\gamma}~,
\end{equation}
and forms an irreducible representation of the group $(\mathbb{Z}^{\widehat N_A}/\widehat K_A\mathbb{Z}^{\widehat N_A})\times\mathbb{Z}_2$. In addition, we can use the $\widehat K_A$ matrix to compute the braiding of lines
\begin{equation}\label{Sdef}
S(\vec\alpha,\vec\beta)={1\over\sqrt{|\det \widehat K_A}|}\exp\left(2\pi i\vec\alpha^T\widehat K_A^{-1}\vec\beta\right)~,
\end{equation}
and their self-statistics / topological spin
\begin{equation}\label{Tdef}
\theta(\vec\alpha)=\exp\left(\pi i\vec\alpha^T\widehat K_A^{-1}\vec\alpha\right)~.
\end{equation}
Note that the topological spin does not depend on the choice of element in the equivalence class. To understand this statement, consider
\bea
\theta(\vec\alpha+\widehat K_A\vec\gamma )&=& \exp\Big( \pi i  (\vec \alpha+\widehat K_A\vec\gamma)^T \widehat K_A^{-1}(\vec\alpha+\widehat K_A\vec\gamma) \Big) =\exp\Big( \pi i  \vec\alpha^T \widehat K_A^{-1}\vec \alpha+ \pi i\vec\gamma^T \widehat K_A \vec\gamma+2 \pi i\vec \gamma^T   \vec \alpha  \Big)\cr&=&\exp\Big( \pi i  \vec\alpha^T \widehat K_A^{-1}\vec\alpha+ \pi i \sum_i (\widehat{K}_A)_{ii}  ( \gamma_i)^2   \Big) ~.
\eea
Due to the condition \eqref{GammaConstraint} and the fact that $\gamma_i-\gamma_i^2\in2\mathbb{Z}$, we have $ \sum_ i\widehat K_{A}(\gamma_i)^2\in2\mathbb{Z}$. Therefore
\be
\theta(\vec\alpha+\widehat K_A\vec\gamma ) =    \theta(\vec \alpha  )~,\qquad  \sum_ i(\widehat K_A)_{ii}\gamma_i\in2\mathbb{Z}~,
\ee
as claimed.

From \eqref{Tdef} and the Abelian fusion rules, we can construct any observable for the TQFT built from the lines (e.g., including the fusion rules in \eqref{FusionRules} via the Verlinde formula). As a final aside, note that these quantities also show that extending $K_A\to\widehat K_A$ does not tensor in any new lines with non-trivial braiding.

Let us now consider the mirror theory and ask what the \lq\lq dual" to the UV ancestor of the universal deformation is. To answer this question, we replace the hypermultiplets and gauge fields in \eqref{hyperUnivGen} and \eqref{vectUnivGen} with mirror hypermultiplets and gauge fields. Since mirror symmetry exchanges $\mathfrak{su}(2)_C\leftrightarrow \mathfrak{su}(2)_H$, and recalling that the algebra in \eqref{univDef} is odd under this exchange, we should take
\begin{equation}
m_B\sim-m_A~.
\end{equation}
In particular, the masses should have opposite sign, because this deformation controls the tree-level masses of $\mathfrak{su}(2)_H$-charged primaries in the mirror theory.\footnote{The sign flip also has interesting consequences. In particular, when the SCFTs are self-mirror symmetric, the resulting IR TQFTs turn  out to be time-reversal invariant \cite{jhltqft}.} Proceeding as in the discussion leading to \eqref{A1loop}, we have 
\begin{equation}\label{B1loop}
(K_B)_{mn}=-{\rm sign}(m_A)\sum_{i=1}^{N_f} \tilde q_{mi}\tilde q_{ni}\ \Rightarrow\ K_B=-{\rm sign}(m_A)\tilde q\tilde q^T~.
\end{equation}
In general, we also should enlarge the $K_B$ matrix as follows
\begin{equation}\label{B1loopExt}
\widehat{K}_B=K_B\oplus \Pi_B~,
\end{equation}
where $\Pi_B$ is a diagonal matrix with entries $\pm1$ that corresponds to stacking the effective theory with a $U(1)_1^{n_B}\boxtimes U(1)_{-1}^{\ell_B}$ SPT. Clearly, $\Pi_{A,B}$ should be chosen so that the IR TQFTs on both sides of the mirror duality have the same central charge (i.e., their signatures should match) and are both spin TQFTs. In the deep IR, we find a mirror TQFT described by
\begin{equation}\label{TQFTB}
S_{IR} = {i\over4\pi}\int d^3x\vec{\tilde a}^T\widehat K_B d\vec{\tilde a}~,
\end{equation}
where $\vec{\tilde a}$ is the vector of dual gauge fields. We can then repeat the discussion in \eqref{lineOps}, \eqref{LatticeID}, \eqref{GammaConstraint}, \eqref{FusionRules}, \eqref{Sdef}, and \eqref{Tdef} with $\widehat K_A$ replaced by $\widehat K_B$.

According to our general discussion, all observables in the TQFTs described by \eqref{TQFTA} and \eqref{TQFTB} should match. This is a highly non-trivial and systematic check of the mirror symmetry duality between the original gauge theories and of our phase diagram.

In the next section, we will discuss the case of arbitrary Abelian gauge theories with unimodular charge matrices (i.e., ${\rm det}(\mathbf{q})=\pm1$) and show that the dual TQFTs do indeed match. Then, in the subsequent section, we will relax this condition. In particular, we will consider theories where we discretely gauge Coulomb branch (topological) symmetries of $\CT_A$ and matter symmetries of the mirror theory, $\CT_B$.\footnote{One can also consider theories in which we discretely gauge both Coulomb and Higgs branch symmetries. For simplicity, we do not carry out this analysis here.}

\newsec{TQFTs from universal-mass-deformed Abelian gauge theories without 1-form symmetry}\label{no1form}
We begin our discussion with the simplest class of $\CN=4$ Abelian gauge theories without 1-form symmetry: $U(1)$ SQED with $N_f$ flavors of charge $+1$.\footnote{More generally, we can consider $U(1)$ SQED with hypermultiplets of co-prime charges and obtain a theory without 1-form symmetry. This example also shows that the condition ${\rm det}(\mathbf{q})=\pm1$ is sufficient but not necessary for the absence of 1-form symmetry (e.g., take $N_f=2$ and assign $U(1)$ flavor charges $+1$ and $-1$ to gauge charge $2$ and gauge charge $3$ hypermultiplets respectively). However, we will argue that any $U(1)^{N_c}$ gauge theory with integer gauge charged hypermultiplets and without 1-form symmetry can be completed to a theory with unimodular $\mathbf{q}$ (in the example of this footnote, instead assign flavor charges $+1$ to both hypermultiplets).} For $N_f>1$, these theories are believed to flow to interacting SCFTs, while, for $N_f=1$, the theory is believed to flow to a free twisted hypermultiplet.

By our discussion in Section \ref{generalDef}, to understand the universal deformation at the IR fixed point, we instead deform the UV SQED theory by adding large mass, $|m_A|\gg g^2$, for the matter fields. In this case, the gauge charge matrix is given by $q_{1i}=1$, and the one-loop $K$-matrix we find via \eqref{A1loop} and \eqref{A1loopExt} is therefore
\begin{equation}\label{U1IR}
\widehat K_A = {\rm sign}(m_A)N_f\oplus\Pi_A~,
\end{equation}
with $\Pi_A$ a diagonal matrix of $\pm1$ entries. In the deep IR, we then find the spin TQFT, $\widehat\CT_A$, given by 
\begin{equation}\label{U1IRTQFT}
S={i\over4\pi}\int d^3x\vec a^T\widehat K_Ad\vec{a}~.
\end{equation}
At the level of line operators, this theory is equivalent to $U(1)_{{\rm sign}(m_A)N_f}$ spin CS theory for odd $N_f$ and $U(1)_{{\rm sign}(m_A)N_f}\boxtimes \langle1,f\rangle$ for even $N_f$ (this is again a spin TQFT, although the first factor in the stacking is a non-spin CS theory). Here $\langle1,f\rangle$ is a theory whose lines are generated by a transparent fermion, $f$, satisfying $f\times f=1$ is the trivial bosonic line.

\begin{figure}
\centering
\begin{tikzpicture}[everynode/.style={circle,draw}]
    \draw ([xshift=-2pt,yshift=-4pt]0,0) rectangle ++(9pt,9pt);
    \draw (1,0) circle (5pt);
    \draw (2,0) circle (5pt);
    \draw (3,0) circle (5pt);
    \draw ([xshift=-2pt,yshift=-4pt]4,0) rectangle ++(9pt,9pt);
    \draw (0.25,0)--(0.82,0);
    \draw (1.21,0)--(1.83,0);
    \draw [dashed] (2.23,0)--(2.83,0);
    \draw (3.2,0)--(3.9,0);
    \draw [decorate,decoration={brace,amplitude=5pt,mirror,raise=4ex}]
  (0.8,0) -- (3,0) node[midway,yshift=-3em]{$N_f-1$};
\end{tikzpicture}
\caption{The quiver for the mirror dual of 3d $\CN=4$ SQED with $N_f$ flavors. It is a $U(1)^{N_f-1}$ quiver gauge theory with (bi)fundamental hypermultiplets emanating from the gauge nodes.}
\label{SQEDBquiv}
\end{figure}
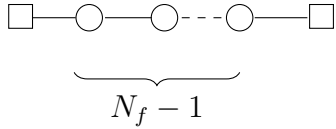

Next, let us consider the mirror theory described in Fig. \ref{SQEDBquiv}. It consists of a $U(1)^{N_f-1}$ theory with $N_f$ hypermultiplets. The gauge charge matrix in this case is
\begin{equation}
q_{aj}=\delta_{aj}-\delta_{a(j+1)}~,
\end{equation}
and the corresponding $K$ matrix is
\begin{equation}
\widehat K_B = -{\rm sign}(m_A)F_{SU(N_f)}\oplus\Pi_B~,
\end{equation}
where $F_{SU(N_f)}$ is the Cartan matrix of $SU(N_f)$. Therefore, in the deep IR we find the TQFT
\begin{equation}
S={i\over4\pi}\int d^3x\vec a^T\widehat K_Bd\vec{a}~.
\end{equation}
At the level of line operators, this theory is equivalent to $SU(N)_{-{\rm sign}(m_A)}\boxtimes\langle1,f\rangle$. This TQFT is again a spin CS theory (although the first factor is a non-spin CS theory) and, it is well-known to be dual to the TQFT in \eqref{U1IR}. This \lq\lq level-rank" duality is often written in a rather imprecise form as
\begin{equation}
SU(N_f)_{\pm1}\leftrightarrow U(1)_{\mp N_f}~,
\end{equation}
but should in fact be supplemented by transparent fermion lines in order for the duality to hold for general $N_f$ (see \cite{Hsin:2016blu} for a more precise statement with a larger set of background fields turned on).

Let us make a few comments:
\begin{itemize}
\item It is instructive to consider the simple case of $N_f=1$ which flows to a single twisted hypermultiplet in the deep IR (these degrees of freedom correspond to a BPS monopole operator of scaling dimension $1/2$ transforming as the ${\bf(2,1)}$ irrep of the $R$ symmetry). The dual theory is a free hypermultiplet (which also has scaling dimension $1/2$ but transforms in the ${\bf(1,2)}$ irrep of the $R$ symmetry). From the discussion in Appendix \ref{Hyperexample}, one can check that the universal deformation of the twisted hypermultiplet with mass $m$ gives the same algebra as that arising from turning on a universal mass $-m$ for an untwisted hyper. Therefore, the IR universal mass deformation has opposite sign from its UV ancestor (and the same sign as the mirror deformation).
\item In our proposal for the UV ancestor of the universal deformation in Section \ref{generalDef}, we set the CS term to zero. Suppose instead we turn on a UV CS term, $k\ne0$, in the SQED theory. Then, in the $N_f=1$ example, we see that the phase diagram must be more complicated for generic $k$. Indeed, in this case, $K_A=k+{\rm sign}(m_A)$. Therefore, for generic $k$, either the IR TQFT contains anyons with non-trivial braiding, or there is no IR CS term at all.\footnote{Consider $m_A>0$. Then, we find an IR theory with trivial braiding if and only if $k=0$ or $k=-2$ (similarly, if $m_A<0$, we find a theory with trivial braiding if and only if $k=0$ or $k=2$). The case with $k\ne0$ is unnatural for the reasons explained in the previous section (it is an additional discrete deformation of the UV theory, and we expect the UV ancestor of the universal mass to correspond to a single continuous deformation).} In the latter case, we have a massless gauge field. On the other hand, turning on the universal mass in the free hypermultiplet gives mass to all the degrees of freedom, and it cannot generate anyons with non-trivial braiding.
\item More generally, our above discussion suggests that the $\CN=4$ gauge coupling / universal mass phase diagram takes a particularly simple form (see Fig. \ref{AbPhase}): we do not expect any phase transitions as we vary $m_A$ relative to the gauge coupling (and similarly for $m_B$ relative to the dual gauge couplings), because the gapped phases from turning on the UV ancestors of the universal mass deformations in each mirror frame give TQFTs that match (and mirror symmetry is an IR duality).
\end{itemize}

In the next section, we add some additional details to our IR descriptions above using the fact that SQED with $N_f>1$ has mixed 't Hooft anomalies for its Coulomb branch and Higgs branch symmetries \cite{Bhardwaj:2022dyt,Bhardwaj:2023zix}. As we will see, this discussion also allows us to  say something about more general abstract local 3d $\CN=4$ SCFTs and their phase diagrams.

\subsec{Anomaly matching in the IR TQFT and Symmetry Fractionalization}\label{AnomMatchSec}
As we have mentioned above, one interesting fact about $\CN=4$ SQED with $N_f>1$ flavors of charge $1$ is that there is a mixed 't Hooft anomaly involving the $\CF_C\cong U(1)$ (Coulomb branch) topological symmetry and the $\CF_H\cong PSU(N_f)$ (Higgs branch) flavor symmetry (this latter symmetry is the one that acts on gauge-invariant operators built out of the hypermultiplets).\footnote{In \cite{Bhardwaj:2023zix}, Coulomb branch and Higgs branch symmetries are more generally referred to as \lq\lq type A" and \lq\lq type B" symmetries respectively.} The anomaly is characterized by the 4d SPT phase \cite{Bhardwaj:2022dyt,Bhardwaj:2023zix}
\begin{equation}\label{A4SQED}
\CA_4=\exp\left({2\pi i\over N_f}\int(c_1^C\ {\rm mod}\ N_f)\ \cup\ \omega_2^H\right)~,
\end{equation}
where $c_1^C$ is the first Chern class of the $U(1)$ bundle, and $\omega_2^H$ is the second Stiefel-Whitney class that parametrizes obstructions of lifting a $PSU(N_f)$ bundle to an $SU(N_f)$ bundle.\footnote{Recall that we have a short exact sequence
\begin{equation}\label{Hcover}
1\longrightarrow\mathbb{Z}_{N_f}\longrightarrow SU(N_f)\longrightarrow PSU(N_f)\longrightarrow 1~.
\end{equation}}

To arrive at the anomaly in \eqref{A4SQED}, we can consider the \lq\lq structure" group formed by the $\CG\cong U(1)$ gauge group and the universal cover of the Higgs branch flavor symmetry group, $F_H\cong SU(N_f)$ \cite{Bhardwaj:2022dyt,Bhardwaj:2023zix}
\begin{equation}
\CS\cong{U(1)\times SU(N_f)\over\CE\cong\mathbb{Z}_{N_f}}~,
\end{equation}
where $\CE\cong\langle h\rangle$ is generated by the diagonal element
\begin{equation}
h=(\exp(2\pi i/N_f),\exp(2\pi i/N_f)\mathds{1}_{N_f})~.
\end{equation}
Then, the mixed $U(1)\times PSU(N_f)$ anomaly can be detected via gauge-$PSU(N_f)$ vortex lines that end on associated \lq\lq fractional" $1/4$-BPS monopole operators and correspond to co-characters
\begin{equation}
\phi: U(1)\to\CS~,
\end{equation}
that do not lift to co-characters $\phi':U(1)\to U(1)\times SU(N_f)$. The $\phi$ co-characters take the form \cite{Bhardwaj:2023zix}
\begin{equation}\label{coCharFrac}
\phi:\exp(i\theta)\mapsto\left(\exp(im\theta),{\rm diag}\left(\exp(im_1\theta),\cdots,\exp(im_{N_f}\theta)\right)\right)~,\ m=n-{\ell\over N_f}~,\ m_j=n_j-{\ell\over N_f}~,
\end{equation}
where
\begin{equation}
n,n_j\in\mathbb{Z}~, \ \ \ \ell\in\left\{0,1,\cdots,N_f-1\right\}~,\ \ \ \sum_jn_j=\ell~.
\end{equation}
Clearly, when $\ell\ne0$, we find co-characters that do not lift to co-characters of type $\phi'$. The corresponding monopole operators have fractional magnetic charge, $m\in\mathbb{Q}$. This fact means that we must also introduce an $N_f$-fold cover of $\CF_C$, $F_C$, that the fractional monopole operators transform under
\begin{equation}\label{Ccover}
1\longrightarrow \CZ_C\cong\mathbb{Z}_{N_f}\longrightarrow F_C\longrightarrow\CF_C\longrightarrow1~,
\end{equation}
where the full structure group of the theory factorizes as $\tilde\CS\cong\CS\times\CF_C$. The associated obstruction class (of lifting $\CF_C$ bundles to $F_C$ bundles) is $\omega_2^C=c_1^C\ {\rm mod}\ N_f$.

To explicitly detect the anomaly, we note that
\begin{equation}
\int_{D_2}\omega_2^H=\ell\in\CE\cong\mathbb{Z}_{N_f}~,
\end{equation}
where $D_2$ denotes a small disk intersecting the gauge-$PSU(N_f)$ vortex line transversally. The $1/4$-BPS monopole that the vortex line ends on defines a homomorphism
\begin{equation}
\gamma:\CZ_H\to\widehat\CZ_C~,
\end{equation}
(see Fig. 16 of \cite{Bhardwaj:2022dyt}) that corresponds to the mixed 't Hooft anomaly in \eqref{A4SQED}.

Given the non-vanishing 't Hooft anomaly in \eqref{A4SQED}, it is interesting to ask how the IR TQFT we found in \eqref{U1IR} and \eqref{U1IRTQFT} reproduces this anomaly. At an even more basic level, we would like to understand how these continuous symmetries are realized in the rigid IR phase.\footnote{Note that the same phenomenon occurs in fractional quantum Hall systems.}

One way to answer this question is the following (see \cite{Cheng:2022nds,Cheng:2022nji} for further details of the general procedure described below). Since the IR TQFT, $\widehat\CT_A\cong\widehat\CT_B$, is rigid and has a finite number of lines (made up of the $U(1)_{N_f}$ lines and their products with a transparent fermion line when $N_f$ is even), these lines cannot be permuted by the continuous symmetry. Instead, the lines can be acted on by various projective phases arising from the microscopic degrees of freedom carrying fractional charges / projective representations under $U(1)\times PSU(N_f)$ (as discussed around \eqref{coCharFrac}). These projective phases give rise to the 't Hooft anomaly.

When we go to the covering groups described in \eqref{Hcover} and \eqref{Ccover}, we see that $F_C\times F_H\cong U(1)\times SU(N_f)$ no longer acts projectively and hence has a completely trivial and non-anomalous action on the IR TQFT (we will assume that any other obstructions for lifting symmetries in fermionic phases also vanish). As a result, we can gauge the $F_C\times F_H$ symmetry and obtain the theory
\begin{equation}\label{trivGauge}
\widehat\CT_A\boxtimes (F_C)_{-\tilde\sigma_H^C}\boxtimes (F_H)_{-\tilde\sigma_H^H}~,
\end{equation}
where the final two factors in \eqref{trivGauge} are CS theories with the corresponding gauge groups and levels indicated. Now, if the mixed 't Hooft anomalies for $\CF_C\times\CF_H$ had vanished (which we know from \eqref{A4SQED} is not the case), we would have been able to condense a diagonal Abelian one-form symmetry, $K\cong\mathbb{Z}_{N_f}\times\mathbb{Z}_{N_f}$, to obtain the theory gotten from $\widehat\CT_A$ by gauging $\CF_C\times\CF_H$
\begin{equation}\label{condTA}
\widetilde\CT_A:=(\widehat\CT_A\boxtimes (F_C)_{-\tilde\sigma_H^C}\boxtimes (F_H)_{-\tilde\sigma_H^H})/K~,
\end{equation}
where the 1-form symmetry is implemented via homomorphisms
\begin{equation}
v:K\to\CA\le\widehat\CT_A~,\ \ \ s:K\to \CA_C\boxtimes\CA_H\le (F_C)_{-\tilde\sigma_H^C}\boxtimes (F_H)_{-\tilde\sigma_H^H}~.
\end{equation}
The targets of these homomorphisms are described as follows: $\CA$ is a subcategory of lines of $\widehat\CT_A$ while $\CA_C$ and $\CA_H$ are Abelian subcategories of lines in the second and third factors of \eqref{trivGauge} respectively. Given the product structure of the symmetry group, we have the following naturally projected homomorphisms
\begin{equation}
s_C:K|_{(\mathbb{Z}_{N_f},0)}\to(F_C)_{-\tilde\sigma_H^C}~,\ \ \ s_H:K|_{(0,\mathbb{Z}_{N_f})}\to(F_H)_{-\tilde\sigma_H^H}~.
\end{equation}
In the case of the $(F_H)_{-\tilde\sigma_H^H}$ theory, the map $s$ ($s_H$) takes elements of $K$ ($K|_{(0,\mathbb{Z}_{N_f})}$) to Abelian lines corresponding to the center, $\CZ_H\cong\mathbb{Z}_{N_f}$. As we have emphasized above, the $\widetilde\CT_A$ TQFT exists if and only if the original $\CF_C\times\CF_H$ symmetry is non-anomalous (which, in our particular case of interest, is not the case).

Since the symmetries we are gauging are bosonic, the anyons implementing $K$ are bosonic
\begin{equation}\label{bosTConst}
\theta_{v(k_C,k_H)}\theta_{s(k_C,k_H)}=\theta_{v(k_C,k_H)}\theta_{s_C(k_C)}\theta_{s_H(k_H)}=1~,\ \ \ \forall (k_C,k_H)\in K~,
\end{equation}
where $\theta_{v(k_C,k_H)}$ is the topological spin of anyon $v(k_C,k_H)\in\widehat\CT_A$, $\theta_{s_C(k_C)}$ is the topological spin of the anyon $s_C(k_C)\in(F_C)_{-\tilde\sigma_H^C}$, and $\theta_{s_H(k_H)}$ is the topological spin of the anyon $s_H(k_H)\in(F_H)_{-\tilde\sigma_H^H}$. In order to be condensable, these lines must also have trivial mutual braiding
\begin{eqnarray}\label{bosSConst}
M_{v(k_C,k_H)v(k'_C,k'_H)}M_{s(k_C,k_H)s(k'_C,k'_H)}&=&M_{v(k_C,k_H)v(k'_C,k'_H)}M_{s_C(k_C)s_C(k'_C)}M_{s_H(k_H)s_H(k'_H)}\cr&=&1~,\ \ \ \forall (k_C,k_H),(k'_C,k'_H)\in K~,
\end{eqnarray}
where
\begin{equation}
M_{ab}:={S_{ab}\over S_{0b}}~,
\end{equation}
and $S_{ab}$ is the modular $S$ matrix. 

Clearly, some of the conditions on the absence of one-form anomalies in \eqref{bosTConst} or \eqref{bosSConst} must be violated in order for us to find the zero-form anomaly \eqref{A4SQED}. This link between the anomalies is summarized in a 4d SPT relation between \eqref{A4SQED} and an SPT for the one-form symmetry corresponding to $K$ \cite{Cheng:2022nds,Cheng:2022nji}
\begin{equation}\label{SPT01rel}
\CA_4[A_C,A_H]=\exp\left(iS_{\rm K-SPT}(B_C[\mu_C[A_C]],B_H[\mu_H[A_H]])\right)~.
\end{equation}
This expression requires some unpacking. First, $A_C$ ($A_H$) are one-form background gauge fields for $\CF_C$ ($\CF_H$), $B_C$ ($B_H$) are two-form background gauge fields corresponding to the two $\mathbb{Z}_{N_f}$ factors in $K$ (let us call them $K_C$ and $K_H$ so that $K\cong K_C\times K_H$), and $\mu_C$ ($\mu_H$) are elements of $H^2(B\CF_C,K_C)$ ($H^2(B\CF_H,K_H)$) corresponding to the group extensions \eqref{Ccover} and \eqref{Hcover}. Second, the one-form gauge fields $A_C$ ($A_H$) induce pullbacks $\mu_C[A_C]\in H^2(X,K_C)$ ($\mu_H[A_H]\in H^2(X,K_H)$), where $X$ is the three manifold our theory is defined on. These objects are precisely the obstruction classes we discussed from the microscopic UV perspective earlier in this section. Finally, the exponential on the RHS of \eqref{SPT01rel} is given by the action of the $K$-SPT
\begin{equation}\label{ActionSPT}
S_{\rm K-SPT}(B_C[\mu_C[A_C]],B_H[\mu_H[A_H]])=2\pi\int h[v(B_C,B_H)]~,
\end{equation}
where $h:\CA\to U(1)$ is the quadratic form that gives the spin of the lines in $\CA$. 

Let us now study how the clash between the conditions in \eqref{bosTConst} and \eqref{bosSConst} leads us to match \eqref{A4SQED} via \eqref{SPT01rel}. More simply, these conditions can be summarised via the independent relations
\begin{equation}\label{simpleCond}
\theta_{v(1,0)}\theta_{s_C(1)}=1~,\ \ \ \theta_{v(0,1)}\theta_{s_H(1)}=1~,\ \ \ M_{v(1,0)v(0,1)}=1~.
\end{equation}
Here $v(1,0)$ and $v(0,1)$ are the lines in $\widehat\CT_A$ that determine the fractionalization patterns for $\CF_C$ and $\CF_H$ (in \cite{Cheng:2022nds,Cheng:2022nji}, these lines are referred to as \lq\lq visons"). Indeed, lines $a\in\widehat\CT_A$ must carry charges $q_a^C$ and $q_a^H$ satisfying
\begin{equation}
\exp\left(2\pi iq^C_a\right)=M_{av(1,0)}~,\ \ \ \exp\left(2\pi iq^H_a\right)=M_{av(0,1)}~.
\end{equation}
We can write the conditions in \eqref{simpleCond} in terms of Hall responses
\begin{eqnarray}\label{HallConds}
\theta_{v(1,0)}\theta_{s_C(1)}=\exp\left({2\pi i \over2N_f}(n_1^2-N_f\sigma_H^C)\right)&=&1~,\ \ \ \theta_{v(0,1)}\theta_{s_H(1)}=\exp\left({2\pi i \over2N_f}(n_2^2-(N_f-1)\sigma_H^H)\right)=1~,\cr M_{v(1,0)v(0,1)}&=&\exp\left({2\pi i\over N_f}n_1n_2\right)=1~,
\end{eqnarray}
where $\sigma_H^C$ and $\sigma_H^H$ are effective CS terms / Hall responses for $\CF_C$ and $\CF_H$ respectively.\footnote{To arrive at the first two conditions in \eqref{HallConds}, we have chosen particular one-form symmetry generators $s_C(1)$ and $s_H(1)$. For example, consider the $(F_C)_{-\tilde\sigma_H^C}$ CS theory. Since $F_C$ is an $N_f$-fold cover of $\tilde\CF_C$, $\tilde\sigma_H^C=N_f^2\sigma_H^C$. Then, one choice of generator of $\mathbb{Z}_{N_f}$ 1-form symmetry is $x^{N_f\sigma_H^C}$ with topological spin $\theta_{x^{N_f\sigma_H^C}}=\exp(-i\pi\sigma_H^C)$ (clearly, a consistency condition is that $N_f\sigma_H^C\in\mathbb{Z}$). The first equation in \eqref{HallConds} is then the condition that the fractional Hall response is determined by the $\CF_C$ charge of the vison: $q_{v(1,0)}=\sigma_H^C$ mod $\mathbb{Z}$. \label{HallVison}} The conditions in \eqref{HallConds} are tantamount, via \eqref{SPT01rel}, to the vanishing of the following phase for some choices of Hall conductances $\sigma_H^C$ and $\sigma_H^H$ (subject to $\sigma_H^C N_f,\sigma_H^H\in\mathbb{Z}$; see Footnote \ref{HallVison})
\begin{eqnarray}
\exp\Big({2\pi i \over2N_f}(n_1^2-N_f\sigma_H^C)\int\CP(\mu_C[A_C])&+&{2\pi i \over2N_f}(n_2^2-(N_f-1)\sigma_H^H)\int\CP(\mu_H[A_H])\cr&+&{2\pi i\over N_f}n_1n_2\int \mu_C[A_C]\cup\mu_H[A_H]\Big)=1~,
\end{eqnarray}
where $\CP(\cdots)$ is the Pontryagin square. It is easy to check that we can always choose $\sigma_H^C$ and $\sigma_H^H$ such that the first two terms in the above phase vanish (this statement corresponds to the fact that there are no separate $U(1)$ and $PSU(N_f)$ anomalies allowed cohomologically \cite{Cheng:2022nds,Cheng:2022nji}). However, for generic choices of $n_1$ and $n_2$, the third phase is non-vanishing and gives rise to the anomaly 
\begin{eqnarray}
\CA_4=\exp\Big({2\pi i\over N_f}n_1n_2\int \mu_C[A_C]\cup\mu_H[A_H]\Big)~.
\end{eqnarray}
In particular, the UV anomaly in \eqref{A4SQED} can be reproduced from the braiding of $v(1,0)$ and $v(0,1)$ provided that
\begin{equation}\label{AnomMatch}
n_1n_2=1\ \ \ {\rm mod} \ N_f~.
\end{equation}
Note that such a choice of $n_1$ and $n_2$ is always possible. For even $N_f$, we can choose $n_1=1$ and $n_2=1-N_f$, while, for odd $N_f$, we can choose $n_1=N_f+1$ and $n_2=1-N_f$.

Let us make a few comments on the anomaly matching we've just discussed:
\begin{itemize}
\item The choice of $n_1$ and $n_2$ in \eqref{AnomMatch} is somewhat arbitrary. Indeed, since the anomaly is fixed by a braiding phase, it is clear that there is an ambiguity related to the action of the zero-form symmetries in $\widehat\CT_A$. These symmetries preserve the $S$ matrix (and all observables in the theory). For generic $N_f$, we expect most of these symmetries to be accidental since they would otherwise mix different vortex sectors in the microscopic theory.
\item Note that for $N_f$ odd, we cannot choose $n_1=1$ or $n_2=1$. The reason is that the corresponding line has order $2N_f$ instead of $N_f$ (its $N_f^{\rm th}$ power is the transparent fermion).
\item We will use the above results to study discrete gauging of $\widehat\CT_A$ and the UV presence of one-form symmetry in Section \ref{1formResults}. We will also see in Section \ref{general} that a generalization of the above logic, combined with structure theorems for (super) MTCs, can be used to show that, under relatively broad conditions, local 3d $\CN=4$ SCFTs certain mixed type A symmetry / type B symmetry 't Hooft anomalies can always be deformed to gapped theories that include a decoupled Abelian TQFT.
\end{itemize}

\subsec{More general theories}
In the previous section, we showed that turning on the UV ancestors of the universal mass deformations of 3d $\CN=4$ SQED with $N_f$ charge-one hypermultiplets and of its mirror dual resulted in dual TQFTs. We argued that this matching suggests that the phase diagrams of these theories as a function of the gauge coupling(s) and the universal mass each contain a single (second-order) phase transition at $m=0$ and strong gauge coupling(s).

In this section, we wish to discuss more general 3d $\CN=4$ Abelian gauge theories. Like the SQED case considered previously, we will restrict our analysis to theories that have no 1-form symmetry (we will return to cases with 1-form symmetry in Sec. \ref{1formResults}) and only integer gauge and flavor charges.\footnote{As described in the previous section, fractional charges arise from certain vortices / monopoles and the fractionalized degrees of freedom they flow to in the IR.} The absence of 1-form symmetry means that the gauge groups act faithfully on the matter content (i.e., there are no subgroups---continuous or discrete---of the $U(1)^{N_c}$ gauge group that leave the matter fields invariant).

Clearly, any unimodular matrix $\mathbf{q}\in GL(N_f,\mathbb{Z})$ in \eqref{Aside} will furnish an example of a theory without 1-form symmetry and a faithful gauge group action. Moreover, this condition is preserved under mirror symmetry (by \eqref{Bside}) and so the mirror will also lack 1-form symmetry.\footnote{One could imagine emergent 1-form symmetry in the SCFT  these theories  both flow to in the IR (when we turn on the $\CN=4$ gauge coupling). In such a case, it is not a priori clear that the UV one-form symmetry would have to match.}

Let us make a few comments:
\begin{itemize}
\item We can find integer flavor charges (i.e., $\hat q$ is an integer matrix) for any Abelian gauge theory with integer gauge charges (i.e., $q$ is an integer matrix) and no 1-form symmetry such that $\mathbf{q}\in GL(N_f,\mathbb{Z})$. In this sense, our Abelian gauge theories without one-form symmetry are general. 

To derive this statement, suppose there were a gauge theory without one-form symmetry that could not be completed to such a $\mathbf{q}$. By Theorem 1 in \cite{zhan2006completion}, such a $q$ has a Smith normal form, $q_S$, with at least one entry $n>1$
\begin{equation}
q_S=UqV~,
\end{equation}
where $U\in GL(N_c,\mathbb{Z})$ and $V\in GL(N_f,\mathbb{Z})$. As a result, $Uq=q_SV^{-1}$ has a row with all entries a multiple of $n$. Since $U$ is a $GL(N_c,\mathbb{Z})$ redefinition of the gauge fields, we see that the theory has (at least) a $\mathbb{Z}_{n}$ one-form symmetry. This is a contradiction.
\item SQED with charge-one fields is clearly in our class of theories. More explicitly, we have
\begin{eqnarray}\label{SQEDuni}
\mathbf{q}&=&\PBK{ 1 &  1 & 1 & \cdots & 1\\  0 &  -1 & -1 & \cdots &-1\\ 0 & 0 & -1& \cdots &-1\\ \cdots&\cdots&\cdots&\cdots&\cdots\\0&0&0&\cdots&-1}~,\cr\mathbf{\tilde q}&=&\PBK{ 1 &  0 & 0 & \cdots & 0 &0\\  1 &  -1 & 0 & \cdots &0&0\\ 0 & 1 & -1& \cdots &0&0\\ \cdots&\cdots&\cdots&\cdots&\cdots&\cdots\\0&0&0&\cdots&1&-1}~.
\end{eqnarray}
\item Our class of theories includes quiver gauge theories, but it also includes infinitely many non-quiver theories (see Fig. \ref{AbGraph}).
\end{itemize}

Since the data from the charge matrices partially determines the IR TQFTs, let us derive a few useful relations among the charge matrices that follow from the action of mirror symmetry described around \eqref{Aside} and \eqref{Bside} and reproduced here
\begin{equation}\label{Aside2}
\mathbf{q} = \begin{pmatrix}
q \\ \hat q
\end{pmatrix}\to\tilde{\mathbf{q}}=\begin{pmatrix}
\tilde{\hat q} \\ \tilde q
\end{pmatrix}~,\ \ \ \tilde{\mathbf{q}}=\mathbf{q}^{-1,T}~.
\end{equation}
To that end, consider the following quantities
\begin{equation}
\mathbf{q}^T\tilde{\mathbf{q}}=q^T\tilde{\hat q}+\hat q^T\tilde q~, \ \ \ \tilde{\mathbf{q}}\mathbf{q}^T
            = \PBK{ \tilde{\hat q} q^T &  \tilde{\hat q} \hat q^T\\   \tilde q q^T &  \tilde q \hat q^T }~.
\end{equation}
From \eqref{Aside2}, it is easy to see that $\mathbf{q}^T\tilde{\mathbf{q}}=\tilde{\mathbf{q}}\mathbf{q}^T=\mathds{1}_{N_f}$. As a result
\begin{equation}\label{gaugeThConds}
q^T\tilde{\hat q}+\hat q^T\tilde q=\mathds{1}_{N_f}~,\ \ \ \tilde q q^T=0~,\ \ \ \tilde{\hat q} \hat q^T=0~,\ \ \ \tilde{\hat q} q^T=\mathds{1}_{N_c}~,\ \ \ \tilde q\hat q^T=\mathds{1}_{N_f-N_c}~.
\end{equation}

We now take this class of theories, turn on the UV deformations subject to $m_A\sim -m_B$, compute the one-loop $\widehat K_A$ and $\widehat K_B$ matrices via \eqref{A1loopExt} and \eqref{B1loopExt}, and flow to the corresponding IR TQFTs $\widehat\CT_A$ \eqref{TQFTA} and $\widehat\CT_B$ \eqref{TQFTB}. We will prove that $\widehat\CT_A\cong\widehat\CT_B$.

However, it is useful to first describe precisely what we mean by an equivalence between two Abelian spin TQFTs described by $K$ matrices $\widehat K_A$ and $\widehat K_B$. Clearly, we require a topological-spin-preserving homomorphism from the theory described by $\widehat K_A$ to the theory described by $\widehat K_B$. In other words, we require a linear map $f: \widehat\CT_A \to \widehat\CT_B$ that preserves fusion and topological spin
\be
f(a\times b) =f(a) \times f(b)~,\ \ \ \theta(a) =\theta(f(a))~,\ \ \  \forall a \in \widehat\CT_A~.
\ee
In fact, for $\widehat K_A$ and $\widehat K_B$ to give rise to dual theories, we actually want an isomorphism between the corresponding theories.  Therefore, we want homomorphisms $f: \widehat\CT_A \to \widehat\CT_B$  and $f': \widehat\CT_B \to \widehat\CT_A$ such that their compositions $f\circ f'=id_B$ and $f'\circ f=id_A$  are identity maps. 
 
Let us ask what these identity maps look like when lifted to our integral lattice defined by the $K$ matrices. Clearly, such maps must send $[\vec\alpha]\to[\vec\alpha]$. At the level of elements of the equivalence class, we require
\be
\vec\alpha \mapsto  \vec\alpha+ K\vec\gamma~,  \ \ \  \sum_ iK_{ii}\gamma_i\in2\mathbb{Z}~, \ \ \ \forall \vec\alpha \in \bZ^n~,
\ee
where $K=\widehat K_{A,B}$. This map is linear, so  it should be implemented by a matrix, $F$, such that $\vec\gamma =X\vec\alpha$, and
\be
F\vec\alpha =  \vec\alpha+ K\vec\gamma =  \vec\alpha+ KX \vec\alpha~.
\ee
Therefore 
\be\label{Fmap}
F=\mathds{1}+KX~,
\ee
subject to 
\be
\sum_i K_{ii}    ( X\vec\alpha)_i= \sum_{i,j } K_{ii}   X_{ij} \alpha_j\in2\mathbb{Z}~,\ \ \ \forall \alpha \in \bZ^n~.
\ee
This requirement leads to the condition 
\be\label{KXeven}
\sum_{i  } K_{ii}   X_{ij}\in2\mathbb{Z}~, \ \ \ \forall j~.
\ee
It is easy to check that $F$ subject to \eqref{KXeven} preserves topological spins.\footnote{To see this explicitly, note that the topological spins are preserved if
\be
F^T K ^{-1}F -K^{-1} =P , \qquad P_{ii}\in2\mathbb{Z} \quad \forall i~.
\ee
Substituting in our definition of $F$, we have
\be\label{PXT}
(1+X^TK ) K^{-1}(1+ KX) -K^{-1}=X^T KX +X^T+X=P~.
\ee
We need only check that $P'=X^T KX$ has even diagonal entries. To see this, note that
\be
P'_{ii}=\sum_{j,k} X_{ji}X_{ki}K_{jk}=2\sum_{j<k} X_{ji}X_{ki}K_{jk}+ \sum_{j} X_{ji}^2 K_{jj}~.
\ee
Now, since $X^2_{ji}-X_{ji}\in2\mathbb{Z}$, \eqref{KXeven} then implies the result and $P$ has even diagonal entries.} Therefore, $f\circ f'$ and $f'\circ f$ lifted to the integral lattice are described by \eqref{Fmap} with $K=\widehat K_B$ and $K=\widehat K_A$ respectively. 

More generally, we need to lift $f$ and $f'$ themselves to the integral lattice. Let us focus explicitly on $f$ ($f'$ is constructed in an analogous manner). Since we require the map to be linear, it should be implemented by a matrix:              
\be
f: [\vec\alpha_1] \mapsto [\vec\alpha_2]~, \qquad \vec\alpha_1+\widehat K_A \vec\gamma_1\mapsto \vec\alpha_2 +\widehat K_B\vec\gamma_2~.
\ee
Therefore, we have an $\widehat N_B\times \widehat N_A$ matrix $\Gamma$ satisfying
\be
\Gamma(\vec\alpha_1+\widehat K_A \vec\gamma_1)= \vec\alpha_2+\widehat K_B\vec\gamma_2~.
\ee
Similarly, for $f'$ we have an $\widehat N_A\times \widehat N_B$ matrix $\Gamma'$ implementing
\be
\Gamma'(\vec\alpha_2'+\widehat K_B \vec\gamma_2')= \vec\alpha_1'+\widehat K_A\vec\gamma_1'~.
\ee
Their compositions are identity maps with lifts given by
\be
\Gamma'\Gamma =\mathds{1}+\widehat K_AY~, \ \ \ \Gamma  \Gamma'=\mathds{1}+\widehat K_BX~,
\ee
subject to 
\be\label{FM2}
\sum_{i} (\widehat K_B)_{ii}   X_{ij}\in2\mathbb{Z}~, \ \ \  \forall j~, \ \ \ \sum_{i} (\widehat K_A)_{aa}   Y_{ab}\in2\mathbb{Z}~, \ \ \ \forall b~.
\ee
Indeed one can check that
\bea
\Gamma' \Gamma(\vec\alpha_1+\widehat K_A\vec\gamma_1)&=&(\mathds{1}+\widehat K_AX)(\vec\alpha_1+\widehat K_A\vec\gamma_1)
=\vec\alpha_1+\widehat K_A \vec\gamma_1  +\widehat K_A  X(\vec\alpha_1+ \widehat K_A\vec\gamma_1)
\cr&=&\vec\alpha_1+\widehat K_A \vec\gamma_1 '~.
\eea
 If $\vec\gamma_1$ is $\widehat K_A$ even, then $\vec\gamma_1'$ is also $\widehat K_A$ even. Therefore $\Gamma'\Gamma$ is indeed an integral lattice lift of the identity map (similar comments apply to $\Gamma\Gamma'$).

Now, for duality to hold, we must also preserve the self-statistics of lines: $\theta(f'(\vec\alpha_2))=\theta(\vec\alpha_2)$ and $\theta(f(\vec\alpha_1))=\theta(\vec\alpha_1)$. This requirement is equivalent to
\bea\label{dualityidentity}
\Gamma'^T \widehat K_A^{-1} \Gamma' - \widehat K_{B}^{-1}&=& P'~, \ \ \  P'_{ii}\in2\mathbb{Z}~,\ \ \  \forall i~,\cr\Gamma^T \widehat K_B^{-1} \Gamma - \widehat K_{A}^{-1}&=& P~, \ \ \  P_{aa}\in2\mathbb{Z}~,\ \ \  \forall a~.
\eea
In summary, we find the following duality conditions at the level of the integral lattice:      
\begin{enumerate}
\item $\Gamma\Gamma'$ acts as an integral lift of the identity
\be
\Gamma  \Gamma'=\mathds{1}+\widehat K_BX~,  \ \ \  \sum_{i} (\widehat K_B)_{ii}   X_{ij}\in2\mathbb{Z}~, \ \ \ \forall j~.
\ee
\item $\Gamma'\Gamma$ acts as an integral lift of the identity 
\be\label{FM2}
\Gamma'\Gamma =\mathds{1}+\widehat K_AY~,\ \ \ \sum_{i} (\widehat K_A)_{aa}   Y_{ab}\in2\mathbb{Z}~,\ \ \  \forall a~.
\ee
\item $\Gamma'$ preserves topological spins
\be
\Gamma'^T \widehat K_A^{-1} \Gamma' - \widehat K_{B}^{-1}= P'~, \ \ \  P'_{ii}\in2\mathbb{Z}~,\ \ \  \forall i~.
\ee
\item $\Gamma$ preserves topological spins
\be
\Gamma^T \widehat K_B^{-1} \Gamma - \widehat K_{A}^{-1}= P~, \ \ \  P_{aa}\in2\mathbb{Z}~,\ \ \  \forall a~.
\ee
\end{enumerate}

In fact, not all of these statements are independent. For example, (4) follows from (3) and (2). Indeed, from (3) we get
\be 
\Gamma^T\Gamma'^T \widehat K_A^{-1} \Gamma'\Gamma - \Gamma^T\widehat K_{B}^{-1}\Gamma=\Gamma^T P'\Gamma~.
\ee
Using (2), this equality is equivalent to
\bea
\Gamma^T P'\Gamma&=&(1+\widehat K_AX)^T \widehat K_A^{-1}  (1+\widehat K_AX)-\Gamma^T\widehat K_{B}^{-1}\Gamma\cr&=&X^T \widehat K_AX+\widehat K_A^{-1}+X+X^T- \Gamma^T\widehat K_B^{-1}\Gamma~. 
\eea
therefore
\be
\Gamma^T\widehat K_B^{-1}\Gamma -\widehat K_A^{-1}=P~, \ \ \  P=X^T \widehat K_AX+X+X^T- \Gamma^T\widehat P'\Gamma~.
\ee
The diagonal elements of $P$ are
\beqn
P_{ii}&=&2X_{ii}-\sum_{jk} \Gamma_{ji}\Gamma_{ki} P'_{jk}+ \sum_{jk} X_{ji}X_{ki} (\widehat K_{A})_{jk}\\&=&2X_{ii}+2 \sum_{j<k} \Gamma_{ji}\Gamma_{ki} P'_{jk}+ 2\sum_{j<k} X_{ji}X_{ki} (\widehat K_A)_{jk}+ \sum_{ j} \Gamma_{ji}\Gamma_{ji} P'_{jj}+ \sum_{j } X_{ji}X_{ji} (\widehat K_A)_{jj}~.\nonumber
\label{evecond}
\eeqn
Therefore (using the fact that the $P'_{ii}$ are even from (3))
\be
P_{ii} \mod 2=\sum_{j } X_{ji}X_{ji} (\widehat K_{A})_{jj}\mod 2=0~.
\ee    
As a result, the duality conditions reduce to (1), (2), and (3).  Note that (1) cannot be removed, because we can consider the case that $\widehat K_B$ is block diagonal, and one block is given by $\widehat K_A$. Then (2) and (3) can be satisfied, but obviously the duality does not generally hold between the theories described by $\widehat K_A$ and $\widehat K_B$.

It is also easy to see from \eqref{evecond} that the evenness conditions in (1) and (2) can be obtained from the evenness conditions in (3) and (4). Therefore, we can also show duality by proving (1)-(4), where the evenness conditions are only imposed on (3) and (4). We will soon see that this alternate approach will prove simpler.

Our strategy will be to first prove (1)-(4) without the evenness conditions directly from the UV gauge theory data. Stacking with appropriate factors of $U(1)_{\pm1}$ will then give the desired evenness result in (3) and (4) and, by the above discussion, a proof of duality.

To that end, we begin by constructing the following matrices built from the UV data: 
\be
\CP =q^T (qq^T)^{-1}q~, \ \ \ \widetilde \CP=   \tilde q^T (\tilde q\tilde q^T)^{-1}\tilde q~.
\ee
Using \eqref{gaugeThConds}, it is easy to check that these matrices are projectors satisfying
\be\label{projectors}
\CP^2=\CP~, \ \ \  \widetilde \CP^2= \widetilde \CP~, \ \ \   \widetilde \CP \CP=\CP  \widetilde \CP=0~,\ \ \  \CP+ \widetilde \CP=\mathds{1}~.  
\ee
Taking the last equation in \eqref{projectors} and multiplying on the left by $\hat q$ and on the right by $\hat q^T$, we obtain
\be\label{qQidentity}
\Gamma'^TK_A^{-1}\Gamma' - K_B^{-1} = {\rm sign}(m_A)\hat q\hat q^T=P'~,\ \ \ \Gamma'=q\hat q^T~.
\ee 
where we have used \eqref{gaugeThConds} and substituted the expressions for the 1-loop $K$ matrices in \eqref{A1loop} and \eqref{B1loop}. This is duality criterion (3) for the 1-loop $K$ matrices (without the evenness condition).

Now, multiplying the last equation in \eqref{projectors} on the left by $\tilde{\hat q}$ and on the right by $\tilde{\hat q}^T$, we obtain
\be
\Gamma^T K_B^{-1}\Gamma -  K^{-1}_A=-{\rm sign}(m_A)\tilde{\hat q}\tilde{\hat q}^T=P~, \ \ \ \Gamma=\tilde q\tilde{\hat q}^T~.
\ee  
This is duality criterion (4) for the 1-loop $K$ matrices (again, without the evenness condition). Now, we can use the definition of $\Gamma$ and $\Gamma'$ to obtain
\be
\Gamma\Gamma'=\tilde q\tilde{\hat q}^Tq\hat q^T=\tilde q(\mathds{1}_{N_f}-\tilde q^T\hat q)\hat q^T=\mathds{1}_{N_f-N_c}-\tilde q\tilde q^T\hat q\hat q^T=\mathds{1}-K_BP'~,
\ee
where we have used \eqref{gaugeThConds}. This is duality condition (1) for the 1-loop $K$ matrices (again, without the evenness condition). Finally, we derive
\be
\Gamma'\Gamma=q\hat q^T\tilde q\tilde{\hat q}^T=q(\mathds{1}_{N_f}-q^T\tilde{\hat q})\tilde{\hat q}^T=\mathds{1}_{N_f-N_c}-qq^T\tilde{\hat q}\tilde{\hat q}^T=\mathds{1}+K_AP~,
\ee
which is duality condition (2) for the 1-loop $K$ matrices (again, without the evenness condition). Therefore, we have explicitly shown the duality conditions (1)-(4) modulo the evenness conditions.

Now we would like to stack with $U(1)_{\pm1}$ factors and establish the $\widehat\CT_A\cong\widehat\CT_B$ duality by showing that we can satisfy the evenness conditions for (3) and (4) for the extended $K$ matrices \eqref{A1loopExt} and \eqref{B1loopExt} which we rewrite as
\be
\widehat{K}_B =\PBK{    K_B &0 \\ 0 & \Pi_B}, \qquad
\widehat{K}_A =\PBK{   K_A &0 \\ 0 & \Pi_A}~,
\ee
where $\Pi_{A,B}$ are diagonal matrices of $\pm1$ that can be used to match the central charge. In fact, without loss of generality, we can choose $\Pi_{A,B}$ to be proportional to the identity and with $r_A={\rm rank}(\Pi_A)\ge{\rm rank}(\Pi_B)=r_B$, and so we have
\be
\widehat{K}_B =\PBK{ K_B &0 \\ 0 & \mathds{1}_B}~, \qquad
\widehat{K}_A =\PBK{   K_A &0 \\ 0 & \mathds{1}_A}~.
\ee
Now, construct two matrices 
\be
\widehat \Gamma'=\PBK{ \Gamma'& 0 \\ J_A & S_A}~, \ \ \ \widehat\Gamma= \PBK{ \Gamma& 0 \\ J_B & S_B}~.
\ee
Here $S_A$ and $S_B$ are, respectively, $r_A\times r_B$ and $r_B\times r_A$ matrices in Smith normal form with all invariants equal to one. Then, we can easily compute     
\beqn
\widehat \Gamma'^T \widehat{    K}_A^{-1} \widehat \Gamma'&=&\PBK{ \Gamma'^T& J_A^T \\ 0 & S^T_A} \PBK{K_A^{-1} &0 \\ 0 & \mathds{1}_A}\PBK{ \Gamma'& 0 \\ J_A & S_A} =\PBK{ \Gamma'^T& J_A^T \\ 0 & S_A^T} \PBK{   K_A^{-1} \Gamma'&0  \\ J_A & S_A}\nonumber\\ &=& \PBK{ \Gamma'^T K_A^{-1}\Gamma'+ J^T_A J_A& J_A^TS_A \\ S_A^TJ_A & S_A^TS_A}=\PBK{  K_B^{-1} +P' + J^T_A J_A& J^T_AS_A \\ S_A^TJ_A &S_A^TS_A}\\&=&\widehat{K}_B^{-1}+\PBK{ P' + J_A^T J_A& J_A^TS_A \\ S_A^TJ_A & 0}=\widehat K_B^{-1}+\widehat P'\nonumber~.
\eeqn
Now, $(  \widehat P')_{ii}=P'_{ii}+(J_A^TJ_A)_{ii}$, and $(\widehat P')_{ii}=0$ for the remaining $r_B$ entries.  Clearly $\widehat P'$ can be brought to have only even diagonals. Similar statements hold under the interchange of $\widehat K_A\leftrightarrow\widehat K_B$ and $\widehat\Gamma'\leftrightarrow\widehat\Gamma$. Therefore, by stacking an SPT we have proven that (3) and (4) hold including the relevant evenness conditions.

Finally, we can also show that
\beqn 
\widehat \Gamma' \widehat\Gamma&=&\PBK{ \Gamma'& 0 \\ J_A & S_A}\PBK{ \Gamma& 0 \\ J_B & S_B}=\PBK{ \Gamma'\Gamma& 0 \\ J_A\Gamma+S_AJ_B & S_AS_B}=\PBK{ \mathds{1}+K_AP& 0 \\ J_A\Gamma+S_AJ_B & S_AS_B}\cr&=&\PBK{ \mathds{1}+K_AP& 0 \\ J_A\Gamma+S_AJ_B & S_AS_B}=\mathds{1}+\PBK{K_AP& 0 \\ J_A\Gamma+S_AJ_B & S_AS_B-\mathds{1}}\cr&=&\mathds{1}+\PBK{K_A& 0 \\ 0 & \mathds{1}_A}\PBK{P& 0 \\ J_A\Gamma+S_AJ_B & S_AS_B-\mathds{1}}=\mathds{1}+\widehat K_A\widehat P~.
\eeqn 
The matrix $S_AS_B-\mathds{1}$ requires some explanation. If $r_A=r_B$, this matrix vanishes. But, in principle, we can have $r_A>r_B$ (recall that, without loss of generality, we assumed that $r_A\ge r_B$). In this case, the $r_A\times r_A$ matrix has vanishing first $r_B$ diagonal entries with the remaining $r_A-r_B$ diagonal entries equal to $-1$. Therefore, we establish (2) without the evenness condition. We can similarly establish (1) without the evenness condition. However, as we discussed previously, since we have established (3) and (4) with the evenness condition, (1) and (2) also must hold with the evenness condition. As a result, we have proven the duality.

Let us make some comments:
\begin{itemize}
\item We have established that, in any 3d $\CN=4$ Abelian gauge theory (which we argued we can always complete to a theory with gauge / flavor charge matrix in $GL(N_f,\mathbb{Z})$), we have a duality of TQFTs when we turn on the UV ancestor of the universal mass deformation. This argument strongly suggests that the phase diagram of all Abelian theories without one-form symmetry (with charges completed in the way we have described) takes the simple form in Fig. \ref{AbPhase}.
\item It seems likely that from our one-loop integrating out of matter, we can produce any positive or negative semidefinite $K$ matrices (physically this is because we are allowed to integrate out arbitrarily complicated matter content for fixed number of gauge fields participating in the TQFT). It would be interesting to understand if dualities then let us explore all possible Abelian spin topological phases (perhaps the fact that our flavor/gauge charge matrices are valued in $GL(N,\mathbb{Z})$ will play a role).
\item In the next section, we will discretely gauge subgroups of the topological Coulomb branch symmetry to produce theories with 1-form symmetry and extend our results further.
\end{itemize}         

\newsec{Universal masses and TQFTs from Abelian gauge theories with 1-form symmetry}\label{1formResults}
In this section, we study 3d $\CN=4$ Abelian gauge theories with one-form symmetry. Our strategy (building on~\cite{Bhardwaj:2023zix}) is to gauge discrete subgroups of the topological / Coulomb branch symmetry appearing in the previous section. In other words, we gauge
\begin{equation}
\CG_T\cong\mathbb{Z}_{n_1}\times\mathbb{Z}_{n_2}\times\cdots\times\mathbb{Z}_{n_{N_c}}<\CF_C^{N_c}\cong U(1)^{N_c}~,
\end{equation}
where the global symmetries on the righthand side are those generated by the $\star F_a$ currents corresponding to each of the $U(1)$ gauge groups ($a=1,\cdots,N_c$ runs over the gauge groups).

Gauging $\CG_T$ produces a dual 1-form symmetry
\begin{equation}\label{1formSymm}
\widetilde\CG_1\cong\mathbb{Z}_{n_1}\times\mathbb{Z}_{n_2}\times\cdots\times\mathbb{Z}_{n_{N_c}}~.
\end{equation}
Note that $\widetilde\CG_1$ will often form a Higgs branch / type $B$ 2-group symmetry with (part of) the Higgs branch $0$-form symmetry \cite{Bhardwaj:2022dyt,Bhardwaj:2023zix}.\footnote{For example, a 2-group can be generated when we gauge a discrete subgroup of a 0-form symmetry that participates in a  mixed anomaly with another 0-form symmetry (which may include an emergent symmetry in the deep IR; these emergent symmetries should have a consistent description in the resulting IR TQFT after turning on the universal deformation). See also the discussion in \cite{Benini_2019}. \label{2groupFootnote}} We can re-interpret the resulting theories as arising from taking the gauge charges in each corresponding row of a $\mathbf{q}$ appearing in the previous section and multiplying by the corresponding $n_a$\footnote{This discussion follows from the arguments in \cite{Gaiotto:2008ak,Borokhov:2002cg}.}
\begin{equation}
q_{ai}\to n_aq_{ai}~.
\end{equation}
Since our original class of theories included the most general $q$ matrices with Smith normal form consisting of all invariants equal to $1$, we see that the class of theories we are considering here consists of the most general $q$ matrices with arbitrary Smith normal form. In other words, the class of theories we consider here are the most general integer charge 3d $\CN=4$ Abelian gauge theories with compact gauge group and standard supermultiplets.\footnote{The 1-form symmetries (and 2-groups) in quiver theories of this type have been studied in \cite{Nawata:2023rdx,Bhardwaj:2023zix}.} These theories are mirror-dual to ones in which we couple finite Abelian Dijkgraaf-Witten theories to dual (Higgs branch) flavor symmetries (see \cite{Kapustin_2014,Balasubramanian:2024nei} for a related discussion).\footnote{Wilson lines in these theories that end on matter fields furnish the one-form symmetry of the dual theory, while the remaining lines are rendered non-topological. Therefore, unlike in the previous section, both mirror dual pairs have UV one-form symmetry. In the flow to the IR SCFT (i.e., without turning on the universal mass), our results suggest that there is no emergent one-form symmetry. Of course, after turning on the universal mass, there is emergent Abelian one-form symmetry (in the next section we will comment on the emergence of such one-form symmetries in more general SCFTs).}

The main idea behind our analysis is the following. Discretely gauging symmetries of the RG flows in the previous section does not change the local dynamics of the QFTs in question. Therefore, we do not expect this procedure to change the number of second-order phase transitions. The main difference is that this process can (and does) change the nature of the IR TQFTs (along with global properties of the SCFTs governing the second-order phase transitions).\footnote{Intuitively this statement is clear, because we are including one-form symmetry lines directly in the UV.} However, if we discretely gauge dual TQFTs, we still have a duality between the resulting theories. Therefore, we do not expect this procedure to change the nature of the phase diagram (beyond changing topological properties of the resulting theories). To illustrate this intuition explicitly, we implement the discrete gauging procedure on both sides of the SQED mirror duality discussed in the previous section. 

\subsec{Explicit computation in SQED and its mirror}
In this section, we explicitly consider 3d $\CN=4$ SQED with $N_f$ hypermultiplets of charge $1$ and its mirror. To that end, recall from \eqref{SQEDuni} that the relevant charge matrices are
\begin{eqnarray}\label{SQEDuni2}
\mathbf{q}&=&\PBK{ 1 &  1 & 1 & \cdots & 1\\  0 &  -1 & -1 & \cdots &-1\\ 0 & 0 & -1& \cdots &-1\\ \cdots&\cdots&\cdots&\cdots&\cdots\\0&0&0&\cdots&-1}~,\cr\mathbf{\tilde q}&=&\PBK{ 1 &  0 & 0 & \cdots & 0 &0\\  1 &  -1 & 0 & \cdots &0&0\\ 0 & 1 & -1& \cdots &0&0\\ \cdots&\cdots&\cdots&\cdots&\cdots&\cdots\\0&0&0&\cdots&1&-1}~.
\end{eqnarray}

Let us now gauge a discrete $\mathbb{Z}_n<\CF_C$ subgroup of the SQED topological symmetry. As follows from \eqref{1formSymm}, the resulting theory has a $\widetilde\CG_1\cong\mathbb{Z}_n$ 1-form symmetry. This QFT can be reinterpreted as SQED with $N_f$ hypermultiplets of charge $n>1$. If we deform the UV theory with large universal mass, $m_A$, for the matter fields (as in the case of Sec. \ref{no1form} for $n=1$), we find a non-trivial extension of the $n=1$ TQFT
\begin{equation}\label{u1qq}
\widehat\CT_A\cong U(1)_{{\rm sign}(m_A)N_fn^2}\boxtimes{\rm SVec}~.
\end{equation}

What happens in the mirror theory? Here we are gauging a $\mathbb{Z}_n$ subgroup of the $U(1)$ flavor (Higgs branch) symmetry. This procedure amounts to coupling a $\mathbb{Z}_n$ Dijkgraaf-Witten theory to the hypermultiplet corresponding to the first column of $\tilde{\mathbf{q}}$ (see \eqref{SQEDuni2}). This hypermultiplet transforms electrically with $\mathbb{Z}_n$ charge $+1$.\footnote{Therefore Wilson lines of $\mathbb{Z}_n$ can end on charged fields, and only the Wilson lines of the Dijkgraaf-Witten theory remain topological in the UV.} As a result, when we integrate out the matter fields, our $(N_f-1)\times (N_f-1)$ $K_B$ matrix in the case of $n=1$ is extended to the following $(N_f+1)\times (N_f+1)$ matrix
\begin{equation}\label{KBSQEDn}
K_B=-{\rm sign}(m_A)\PBK{ 0 & n & 0 & 0 & 0 & \cdots & 0 &0\\  n &  1 & 1 & 0 &0 &\cdots &0&0\\ 0 & 1 & 2& -1 &0& \cdots &0&0\\ 0 & 0 & -1& 2 &-1& \cdots &0&0\\ \cdots&\cdots&\cdots&\cdots&\cdots&\cdots&\cdots&\cdots\\0&0&0&0&0&0\cdots&-1&2}~.
\end{equation}
The upper left $2\times2$ matrix is the $K$ matrix for the $\omega=1\in H^3(\mathbb{Z}_n,U(1))$ twisted $\mathbb{Z}_n$ Dijkgraaf-Witten theory. The twist comes from integrating out the $\mathbb{Z}_n$-charged hypermultiplet (as can be seen from a one-loop Feynman diagram computation with electric Dijkgraaf-Witten external gauge fields; see also \cite{Cordova:2017vab}). The $(2,3)$ and $(3,2)$ entries arise from 1-loop diagrams with one external electric gauge field and the $a=2$ gauge field. The rest of the $K_B$ matrix is identical to that of the $n=1$ theory, because the Dijkgraaf-Witten theory does not couple to matter charged under other parts of the gauge group.\footnote{This matrix is of course not unique. For example, if we start from the different UV $SL(N_f,\mathbb{Z})$ $\tilde{\mathbf{q}}$ matrix in (3.31) of \cite{Bullimore:2015lsa}, we find
\begin{equation}
K_B=-{\rm sign}(m_A)\PBK{ 0 & n & 0 & 0 & 0 & \cdots & 0 &0\\  n &  1 & -1 & -1 &-1 &\cdots &-1&-1\\ 0 & -1 & 2& 1 &1& \cdots &1&1\\ 0 & -1 & 1& 2 &1& \cdots &1&1\\ \cdots&\cdots&\cdots&\cdots&\cdots&\cdots&\cdots&\cdots\\0&-1&1&1&1&\cdots&1&2}~,
\end{equation}
which is dual to \eqref{KBSQEDn}.
}

It is straightforward to check that
\begin{equation}
\widehat\CT_A\cong\widehat\CT_B~,
\end{equation}
where we have extended the $K_A$ and $K_B$ matrices to $\widehat K_A$ and $\widehat K_B$ respectively. This statement is non-trivial evidence that our general expectation is correct: the discrete gauging of the theories in Sec. \ref{no1form} only changes the IR TQFTs via various extensions (and does not add new second-order phase transitions). The conjectured phase diagram takes the form shown in Fig. \ref{AbPhase}.

\newsec{General results on universal deformations of local 3d $\CN=4$ SCFTs}\label{general}
One interesting (and obviously very difficult) open question regarding (S)CFTs is to understand the conditions under which they can be connected to Lagrangian QFTs via an RG flow. One often thinks about this question from the perspective of starting with some weakly coupled fields in the UV and asking if the theory can flow to an (S)CFT in the deep IR (e.g., as in the case of the SCFTs arising in this paper at the IR endpoints of RG flows from Abelian 3d $\CN=4$ gauge theories). But it is also interesting to ask when (S)CFTs can be connected to Lagrangian theories in the deep IR.

This latter phenomenon occurs in SCFTs when one has a moduli space of supersymmetric vacua (see also \cite{Cuomo:2024fuy} for a potentially more general discussion). Indeed, turning on generic expectation values for SCFT operators parameterizing the moduli space often takes one to a Lagrangian theory consisting of the dilaton multiplet for spontaneous breaking of conformal symmetry along with other free fields. More general flows may take one to a decoupled dilaton multiplet stacked with a non-Lagrangian IR SCFT (e.g., as in the case of tensor branch RG flows from the $\mathfrak{su}(N)$ 6d $(2,0)$ SCFT to the $\mathfrak{su}(N-1)$ $(2,0)$ theory when $N>2$).

In this section, we would instead like to start from an abstract local 3d $\CN=4$ SCFT that may or may not have a moduli space of vacua. Instead, we assume that it has a $U(1)\times PSU(N)$ symmetry (which, for simplicity, does not form part of a 2-group) with anomaly
\begin{equation}\label{A4SQEDgen}
\CA_4=\exp\left({2\pi iq\over N}\int(c_1^C\ {\rm mod}\ N)\ \cup\ \omega_2^H\right)~,
\end{equation}
that only depends on the extension class. So as not to complicate matters, we will also assume that {\it$N$ is prime}. We expect extensions of our discussion to hold for more general $N$ (and also for more general symmetries).

Note that the anomaly for SQED described in \eqref{A4SQED} is of the form \eqref{A4SQEDgen} with $N=N_f$ and $q=1$. Here we write $N$ instead of $N_f$ to remind the reader that we have in mind an abstract local 3d $\CN=4$ SCFT that, a priori, may not be related in any way to SQED in the UV (and hence there is no notion of a \lq\lq number of flavors"). Our claim is the following: 

\medskip\medskip\medskip
\noindent
{\bf Claim:} Any local unitary\footnote{As an aside, it is interesting to note that locality (and unitarity) also plays an important role in the representation theoretical analysis of somewhat related phenomena in \cite{Buican:2023efi}.} 3d $\CN=4$ SCFT with a $U(1)\times PSU(N)$ symmetry (where $N$ is prime) and an anomaly of the form \eqref{A4SQEDgen} can be continuously deformed to an IR TQFT that contains an Abelian factor:
\begin{equation}
\widehat\CT_{IR}\cong\CT_{\rm Ab}\boxtimes\CT~,
\end{equation}
where $\CT_{\rm Ab}$ is an Abelian spin TQFT, and $\CT$ is a general spin TQFT over which the anomaly \eqref{A4SQEDgen} trivializes.

\medskip\medskip\medskip
\noindent
Since $\CT_{Ab}$ is an Abelian TQFT, it can be described by an Abelian Chern-Simons theory.\footnote{It is an open question whether all non-Abelian (spin) TQFTs can also be described by Lagrangians.} Hence, we learn that

\medskip\medskip\medskip
\noindent
{\bf Claim$'$:} Any local unitary 3d $\CN=4$ SCFT with a $U(1)\times PSU(N)$ symmetry (where $N$ is prime) and an anomaly of the form \eqref{A4SQEDgen} can be continuously deformed to an IR QFT that contains a decoupled factor described by a Lagrangian. 

\medskip\medskip\medskip
\noindent

In order to establish the above claims, we note that the universal mass deformation, present in any local 3d $\CN=4$ SCFT, preserves all continuous internal symmetries. As we have discussed at great length above, turning it on takes the theory to a gapped phase\cite{Cordova:2016xhm} (if the UV theory is unitary). In particular, the IR is a spin TQFT. Assuming this TQFT is finite (which we believe to be true by unitarity and the $F$-theorem\footnote{Note that topological degrees of freedom contribute to $F$. Therefore, as long as the SCFT has finite $F=F_{UV}$, we conjecture that the IR is always a finite unitary (and, we believe, semisimple) TQFT (it must satisfy $F_{IR}<F_{UV}$ by the $F$-theorem).}), it is described by a super modular tensor category (SMTC). This is a category of lines with non-degenerate braiding except for a trivial line and a transparent fermion; these simple objects (we will use objects and lines interchangeably in what follows) generate a subcategory of transparent lines that is referred to in the category theory literature as \lq\lq ${\rm SVec}$." This ${\rm SVec}$ subcategory is the M\"uger center of the SMTC (the subcategory of objects that have trivial braiding with all objects in the category).

In this setting, we can prove the following theorem that establishes the above claims:

\medskip\medskip\medskip
\noindent
{\bf Theorem:} Let $\mathcal{C}$ be an SMTC with a 't Hooft anomalous global symmetry $G = U(1)\times PSU(N)$ (where $N$ is prime) with the obstruction given as in \eqref{A4SQEDgen}. Then, $\mathcal{C}$ contains a decoupled Abelian spin-TQFT, $\mathcal{B}$, such that $\mathcal{C} = \mathcal{B}\boxtimes_{\rm SVec} C_{\mathcal{C}}(\mathcal{B})$, and $G$ is non-anomalous over $C_{\mathcal{C}}(\mathcal{B})$.

\medskip\medskip\medskip
\noindent
Before giving a proof, let us unpack some of the notation in the statement of the theorem. The notation $X\boxtimes_{\rm sVec}Y$ for SMTCs $X$ and $Y$ means that we take the Deligne (tensor) product of the two theories and condense the algebra $\CA=(1,1)+(f,f)$, where $(1,1)\in X\boxtimes Y$ is the identity line, and $(f,f)\in X\boxtimes Y$ is the product of the transparent fermions in $X$ and $Y$.\footnote{The basic reason for condensing $\CA$ is so that we do not have more than one transparent fermion (and hence so that we do not have a non-trivial transparent boson).} In other words, we have the SMTC
\begin{equation}
X\boxtimes_{\rm sVec}Y\cong(X\boxtimes Y)/\CA~, \ \ \ \CA=(1,1)+(f,f)~.
\end{equation}
In fact, we already made similar (though more telegraphic) comments in footnote \ref{CondenseTensor}. The last notation to unpack in the statement of the theorem is $C_{X}(Y)$ (where $Y\le X$ is a subcategory): it is the subcategory of lines in $X$ that braid trivially with all lines in $Y$.

Given this groundwork, we are ready to prove the theorem.

\bigskip
\noindent
{\bf Proof:} Recall from our discussion around \eqref{SPT01rel} that the anomaly for a continuous global 0-form symmetry acting on a TQFT (in this case the associated SMTC, $\CC$) is realized via a $1$-form anomaly for visons in the pointed (i.e., Abelian) subcategory, $\text{Inv}\big(\mathcal{C}\big)$ (this aspect of our discussion did not depend on the particular TQFT or the associated $\CC$ we were studying in that section). For $G = U(1)\times PSU(N)$, the corresponding obstruction is realized by a homomorphism, $\nu: \zz_N \times \zz_N \rightarrow \text{Inv}(\mathcal{C})$, and visons, $\nu(1,0)$ and $\nu(0,1)$, satisfying the braiding relation
\begin{equation}\label{visBraidq}
M_{\nu(1,0),\nu(0,1)} = \exp\left(\frac{2\pi i q}{N}\right)~.
\end{equation}
In our previous discussion we had $q=1$, but taking more general $q$ satisfying ${\rm gcd}(q,N)=1$ will not affect our discussion drastically (note that, since we assume $N$ is prime, any $q$ will do).

Clearly, \eqref{visBraidq} implies that $\nu(1,0)$ and $\nu(0,1)$ must be Abelian objects of order $N$ (i.e., $\nu(1,0)^N=\nu(0,1)^N=1$). Let us denote the image of $\nu \le \text{Inv}(\mathcal{C})$ as $\text{Im}(\nu)$. Since $\nu$ is a homomorphism, $\text{Im}(\nu)$ is closed and forms a pointed braided fusion subcategory of $\mathcal{C}$, which we denote as $\text{Im}(\nu)\cong\mathcal{B}_0 \le\mathcal{C}$. Note that $\widetilde{\mathcal{B}}\cong\mathcal{B}_0\boxtimes{\rm SVec}$ is also a closed fusion subcategory. Here $\mathcal{C}$ is a super-modular category, and so
\begin{equation}
C_{\CC}(\mathcal{C}) \cong {\rm SVec} \cong \langle1,f\rangle~,
\end{equation}
where $1$ is the trivial object (i.e., the trivial line), and $f$ is the transparent fermion. 

By Theorem 7.9 of \cite{Ng_2022}, to complete our proof we need only show that $\widetilde{\mathcal{B}}$ or some subcategory thereof is super-modular. This statement is equivalent to showing that $C_{\widetilde{\mathcal{B}}}(\widetilde{\mathcal{B}}) = {\rm SVec}$ or showing the same statement for some subcategory of $\widetilde{\mathcal{B}}$.

To that end, let us assume that $\mathcal{B}_0$ has a subcategory, $\mathcal{A}<\CB_0$, of invertible and non-trivial transparent bosons. First, note that this assumption allows us to build $\mathcal{B}_{0}$ as an extension of $\mathcal{D}:=\mathcal{B}_{0}/\mathcal{A}$ by $\mathcal{A}$ (in other words, we can gauge an appropriate 0-form symmetry dual to $\CA$ in $\CD$ to produce $\CB_0$)
\begin{equation}
\begin{tikzcd}
1 \arrow[r] &\mathcal{A} \arrow[r] &\mathcal{B}_0  \arrow[r] &\mathcal{D} \arrow[r] &1~.
\end{tikzcd}
\end{equation}
In fact, this statement also holds for $\mathcal{B}$,
\begin{equation}
\begin{tikzcd}
1 \arrow[r] &\mathcal{A}\boxtimes {\rm SVec} \arrow[r] &\mathcal{B}_0\boxtimes {\rm SVec}  \arrow[r] &\mathcal{D} \arrow[r] &1~.
\end{tikzcd}
\end{equation}

Now, since $\nu$ is a homomorphism from $\zz_{N}\times \zz_{N}$, and the visons have order $N$, $\text{Im}(\nu) = \mathcal{B}_0$ must have $\zz_{N}\times\mathbb{Z}_p$ fusion rules, where $N/p\in\mathbb{Z}$. Therefore, $p=1$ or $p=N$.

Let us first suppose $p=1$. From \eqref{visBraidq}, we see that 
\begin{equation}\label{visBraidqn}
M_{\nu(1,0)^n,\nu(0,1)} = \exp\left(\frac{2\pi i qn}{N}\right)\ne1~,\ \ \  \forall n\in\left\{1,\cdots,N-1\right\}~.
\end{equation}
Clearly, there cannot be a non-trivial transparent boson, $b\in\CB_0$, because otherwise $b=\nu(1,0)^n$ for some $n\ne0\ {\rm mod}\ N$, but the braiding in \eqref{visBraidqn} is always non-trivial. 

Next, suppose $p=N$. If there is a non-trivial transparent boson, $b$, then $b$ must have order $N$ or $N^2$ (since $N$ is prime). Clearly \eqref{visBraidqn} forbids $b$ from having order $N^2$ since then $\CB_0\cong{\rm Rep}(\mathbb{Z}_N\times\mathbb{Z}_N)$ would have trivial braiding. Therefore, $b$ must have order $N$. From \eqref{visBraidqn}, we know that
\begin{equation}\label{visBraidqnb}
M_{b\nu(1,0)^n,\nu(0,1)}=M_{b,\nu(0,1)}M_{\nu(1,0)^n,\nu(0,1)}= \exp\left(\frac{2\pi i qn}{N}\right)\ne1~,
\end{equation}
for all  $n\in\left\{1,\cdots,N-1\right\}$. This logic shows that $b\ne\nu(1,0)^n$ for any $n$. Similarly, we can show that $b\ne\nu(0,1)^n$ for any $n$. As a result, we must have that $b=\nu(1,0)^{n_1}\nu(0,1)^{n_2}$ for $n_{1,2}\ne0\ {\rm mod}\ N$. Therefore
\begin{eqnarray}\label{bpowBraid}
1&=&M_{b,\nu(1,0)^{m_1}\nu(0,1)^{m_2}}=M_{\nu(1,0)^{n_1},\nu(1,0)^{m_1}}M_{\nu(0,1)^{n_2},\nu(0,1)^{m_2}}M_{\nu(1,0)^{n_1},\nu(0,1)^{m_2}}M_{\nu(0,1)^{n_2},\nu(1,0)^{m_1}}\cr&=&M_{\nu(1,0)^{n_1},\nu(1,0)^{m_1}}M_{\nu(0,1)^{n_2},\nu(0,1)^{m_2}}\exp\left({2\pi i q\over N}(n_1m_2+n_2m_1)\right)~,
\end{eqnarray}
for any $m_{1,2}\in\left\{0,\cdots,N\right\}$. Let us set $m_2=0$. Then, we conclude that 
\begin{eqnarray}\label{bpowBraidm20}
1=M_{\nu(1,0)^{n_1},\nu(1,0)^{m_1}}\exp\left({2\pi i n_2m_1q\over N}\right)~.
\end{eqnarray}
But this result implies that the subcategory generated by $\nu(1,0)$, $\CB_{0,\nu(1,0)}$, is modular and $\widetilde{\CB}>\CB\cong\CB_{0,\nu(1,0)}\boxtimes{\rm SVec}$ is supermodular. From Theorem 7.9 of \cite{Ng_2022}, we are done. $\square$

Clearly, it would be interesting to broaden the above analysis to more general groups to see what one can say about connections between Lagrangian theories and general local 3d $\CN=4$ SCFTs.

\newsec{Discussion}
In this paper, we have examined how mirror symmetry descends to TQFT dualities under turning on universal mass deformations of Abelian 3d $\CN=4$ theories (and their corresponding IR SCFTs). Along the way we made contact with the physics of quantum Hall conductance and 't Hooft anomaly matching. Clearly, there are many interesting extensions of what we have discussed here. Among them are the following:

\begin{itemize}
\item It would be interesting to study the universal deformation and the resulting phase diagram in more general QFTs like non-Abelian 3d $\CN=4$ gauge theories and the resulting SCFTs. There, the action of mirror symmetry is more subtle. Another interesting direction would be to study this deformation in the context of Chern-Simons-matter theories (e.g., see \cite{Hosomichi:2008jd}). In this case, the analysis of the universal deformation and the resulting IR TQFT may potentially be simpler, since the gauge degrees of freedom are already approximately topological in the SCFT (although less seems to be known about mirror symmetry in this context).
\item It could be useful to more fully incorporate 2-groups and other more general symmetry structures (e.g., as described in \cite{Buican:2023bzl}) into our analysis (see comments below \eqref{1formSymm} and in footnote \ref{2groupFootnote}).
\item It would be interesting to understand if 2d domain walls separating phases of theories with different signs of the universal deformation encode important universal features of the corresponding SCFTs.
\item There may be more to say about the relationship between the TQFTs studied here and those obtained from topological twisting. As we mentioned in the main text, the former typically lead to semi-simple unitary TQFTs, whereas the latter are typically non-unitary and non semi-simple, except in rank-0 when they are in fact semi-simple~\cite{Gang:2018huc,Gang:2021hrd,Ferrari:2023fez}. The reduction of mirror symmetry to level-rank types of dualities observed here as well as in the twists of families of rank-0 deserves further scrutiny. There is also growing evidence that large classes of MTCs (both unitary and non-unitary) may be recovered from 3d supersymmetric gauge theories, perhaps via the 3d-3d correspondence~(see for instance~\cite{Cho:2020ljj
,Gang:2021hrd,Gang:2024tlp}), a fact that may well be related to our study.
\item Clearly, it would be interesting to generalize Section \ref{general} and understand the relation between abstract local universal 3d $\CN=4$ SCFTs, their 't Hoof anomalies, and IR Abelian spin CS theories more broadly. Some of these abstract studies may also shed light on the symmetry structure of rank 0 SCFTs, where there are no associated moduli spaces.
\item Finally, it would be interesting to understand the degree to which what we have discussed here depends on supersymmetry (at various points in our discussion SUSY was broken). Results in \cite{Seiberg:2016gmd} suggest that many of the phenomena discussed here are much more general (indeed, this might be one interpretation of the simplicity of our phase diagram).
\end{itemize}
We hope to return to some of these questions soon.

\ack{We would like to thank C.~Closset, D.~Delmastro, M.~Hosseini, P.~S.~Hsin, R.~Radhakrishnan, B.~Rayhaun, O.~Uddin, and Z.~Zhong for comments and discussions. M.~Bu. would like to thank the International Centre for Mathematical Sciences and the James Clerk Maxwell Foundation for hospitality and a stimulating environment while this work was being undertaken. M.~Ba., M.~Bu., and Z.~D.'s work was partly supported by the STFC under the grant, \lq\lq Amplitudes, Strings and Duality." A.~B. and M.~Bu.'s work was partly supported by The Royal Society under the grant, \lq\lq Relations, Transformations, and Emergence in Quantum Field Theory." H. J.’s work was supported in part
by the STFC Consolidated Grants ST/T000791/1 and ST/X000575/1. No new data were generated or analyzed in this study.}

\newpage

\begin{appendices}
\newsec{Lagrangian theories with the deformed $\CN=4$ algebra}
In this appendix, we discuss a few examples of Lagrangian theories which are invariant under the 3d $\CN=4$ algebra 
\begin{equation}\label{defAlgApp}
\left\{Q_{\alpha a\dot a},Q_{\beta b\dot b}\right\}=\varepsilon_{ab}\varepsilon_{\dot a\dot b}P_{(\alpha\beta)}+ m\varepsilon_{\alpha\beta}\Big(\varepsilon_{ab}(R_C)_{\dot a\dot b}-\varepsilon_{\dot a\dot b} (R_H)_{ab}\Big)~.
\end{equation}

Before getting into the details, let us first introduce our conventions (mainly following \cite{Dedushenko:2016jxl}). We will work in Euclidean signature and use $\varepsilon$ to raise / lower all $\mathfrak{su}(2)$ indices. For example, 
\be
\psi^\beta=\varepsilon^{\beta\alpha}\psi_\alpha~, \qquad \psi_\beta=\varepsilon_{\beta\alpha}\psi^\alpha~, \qquad \varepsilon^{12}=-\varepsilon_{12}=1~, \qquad
\psi^2=-\psi_1~, \qquad \psi^1= \psi_2~.
\ee
We will sometimes suppress $\mathfrak{su}(2)$ spinor indices using the convention $ \psi \chi =\chi \psi =\psi^\alpha \chi_{\alpha} $.
The gamma matrices  are defined by $(\gamma^\mu)_{\alpha}{}^\beta =(\sigma^\mu)_{\alpha}{}^\beta $ which are the standard Pauli matrices $\sigma^1,\sigma^2,\sigma^3$. For convenience, we also introduce
 \be
\p_{\alpha \beta} :=\p_{  \beta\alpha} =\gamma^\mu_{\alpha\beta}\p_\mu~, \qquad\gamma^\mu_{\alpha\beta}=( \gamma^\mu)_{\alpha }{} ^\kappa \epsilon_{  \beta\kappa}~.
 \ee
More explicitly, the $\gamma^\mu_{\alpha\beta}$ matrices are
\be
\gamma^1_{\alpha\beta}=\left(
\begin{array}{cc}
 -1 & 0 \\
 0 & 1 \\
\end{array}
\right)~, \qquad
\gamma^2_{\alpha\beta}=\left(
\begin{array}{cc}
 i & 0 \\
 0 & i \\
\end{array}
\right) ~,\qquad
\gamma^3_{\alpha\beta}=
\left(
\begin{array}{cc}
 0 & 1 \\
 1 & 0 \\
\end{array}
\right)~.
\ee

\subsec{The free hypermultiplet}\label{Hyperexample}
The simplest example of an SCFT is the free massless hypermultiplet. Turning on the universal deformation in this theory, we find
\be\label{hyperUniv}
S=\int d^3 x \left(\p^\mu \tilde    \rho^a \p_\mu    \rho_a -i \tilde \psi^{\dot a}\gamma^\mu  \p_\mu \psi_{\dot a}+{m^2}  \tilde   \rho^a    \rho_a+mi\tilde\psi^{\dot a}\psi_{\dot a}\right)~.
\ee
It is easy to check that this theory is invariant under the following SUSY transformations (see also \cite{Agarwal:2008pu} for a discussion in momentum basis)
\be\label{basicTrans}
[Q_{\alpha a \dot a},\rho_c]=\varepsilon_{ac}\psi_{\alpha \dot a}~, \qquad \left\{Q_{\beta b \dot b},\psi_{\gamma\dot d}\right\}=\varepsilon_{\dot b\dot d }( i \p_{\beta\gamma}+m i\varepsilon_{\beta\gamma})\rho_b~.
\ee
These transformations give rise to the following on-shell SUSY algebra (which is compatible with \eqref{defAlgApp})
\bea\label{SUSYAlgHyper}
[\left\{Q_{\alpha a \dot a}, Q_{\beta  b \dot b}\right\},\rho_d]&=& \varepsilon_{  a   b}\varepsilon_{\dot a \dot b}P_{\alpha\beta}\rho_d-  m \varepsilon_{  \alpha   \beta}\varepsilon_{\dot a \dot b}[(R_H)_{ab},\rho_d]~,\cr [\left\{Q_{\alpha a \dot a}, Q_{\beta  b \dot b}\right\},\psi_{\dot d}^\gamma]&=& \varepsilon_{  a   b}\varepsilon_{\dot a \dot b}P_{\alpha\beta}\psi_{\dot d}^\gamma+ m \varepsilon_{  \alpha   \beta}\varepsilon_{ab}[(R_C)_{\dot a\dot b},\psi_{\dot d}^\gamma]~,
\eea
where $P_{\alpha\beta}=-i \p_{\alpha\beta}$, and
\be\label{hyperRTrans}
[(R_{H})_{ab},\rho_d] =i(\varepsilon_{ad}\rho_b+\varepsilon_{bd}\rho_a)~, \qquad [(R_C)_{\dot a \dot b},\psi_{\dot d}^\beta]=i(\varepsilon_{\dot a\dot d}\rho_{\dot b}^\beta+\varepsilon_{\dot  b\dot  d}\rho_{\dot a}^\beta)~.
\ee
The supersymmetry algebra and transformation rules for $\tilde \rho$ and $\tilde\psi$ are given analogously. 

Let us explicitly verify the second line of \eqref{SUSYAlgHyper}. To that end, note that the fermionic equations of motion are
\be\label{fmEoM}
(  \p_{\alpha \beta}+m\epsilon_{\alpha \beta})\psi^\beta=\p_{\alpha \beta }\psi^\beta+m\psi_{\alpha  } =0~.  
\ee
Then we have
\be
\epsilon^{  \gamma\alpha}(  \p_{\rho \gamma}- m\epsilon_{\rho \gamma})(  \p_{\alpha \beta}+m\epsilon_{\alpha \beta})\psi^\beta
= \epsilon_{  \rho\beta}( - \p^2+m^2 )\psi^\beta=0~, \quad
\ee
which leads to the on-shell condition $p^2+m^2=0$ in momentum space. Meanwhile,  the  equations of motions for scalars are given by  $( - \p^2+m^2 )\rho_d=0$. This means the scalars and fermions have the same masses, as expected from supersymmetry.

From \eqref{fmEoM},   one can show \be\label{usefulFM}
\p_{\alpha\beta}\psi_{\gamma}+\p_{\gamma\beta}\psi_{\alpha}=2\p_{\alpha\gamma}\psi_{\beta}-m\varepsilon_{\alpha\beta}\psi_{\gamma}-m\varepsilon_{\gamma\beta}\psi_{\alpha}~.
\ee
 
With this simple groundwork, we can derive the second line of \eqref{SUSYAlgHyper}, while the first line is straightforward. From \eqref{basicTrans}, we have
\be 
[Q_{\alpha a \dot a}, \{Q_{\beta b \dot b},\psi_{\gamma\dot d} \}]    =\epsilon_{\dot b\dot d }(i\p_{\beta\gamma}+im\epsilon_{\beta\gamma})  [  Q_{\alpha a \dot a},\rho _b]
        =i\epsilon_{\dot b\dot d }( \p_{\beta\gamma}+m\epsilon_{\beta\gamma})   \epsilon_{ab}\psi_{\alpha \dot a}
  ~,
\ee
and so
\be\label{intExp}
[Q_{\alpha a \dot a}, \{Q_{\beta b \dot b},\psi_{\gamma\dot d} \}] +[Q_{\beta b \dot b}  , \{Q_{\alpha a \dot a} ,\psi_{\gamma\dot d} \}]  =i \epsilon_{ab} \Big[\epsilon_{\dot b\dot d }( \p_{\beta\gamma}+m\epsilon_{\beta\gamma})   \psi_{\alpha \dot a}
 - \alpha   \dot a\leftrightarrow \beta\dot b
  \Big]~.
\ee    
Now, we can use the following formula 
\be
F_{\alpha\beta}=-\frac12 F_\gamma{}^\gamma \varepsilon_{\alpha\beta}+F_{(\alpha\beta)}~, \qquad F_{(\alpha\beta)}=\frac12(F_{\alpha\beta}+F_{ \beta\alpha})~,
\ee  
to simplify the terms in the bracket of \eqref{intExp}:   
\bea
\varepsilon_{\dot b\dot d }  \varepsilon_{\beta\gamma}   \psi_{\alpha \dot a}-\alpha\dot a\leftrightarrow \beta\dot b&=&-\varepsilon_{\dot b\dot a }  \varepsilon_{  \gamma(\beta} \psi_{\alpha) \dot d}-\varepsilon_{  \alpha\beta}\psi_{\gamma (\dot a} \varepsilon_{\dot b)\dot d}~,\cr\varepsilon_{\dot b\dot d }  \p_{\beta\gamma}   \psi_{\alpha \dot a}-\alpha   \dot a\leftrightarrow \beta\dot b&=& \varepsilon_{\dot b\dot a}  \p_{\gamma(\beta }   \psi_{\alpha) \dot d}+\varepsilon_{\alpha\beta }  \p_{ \gamma} {}^\eta  \psi_{\eta( \dot a}\varepsilon_{\dot b)\dot d }\cr&=&\varepsilon_{\dot b\dot a}  \p_{\alpha \beta }   \psi_{\gamma  \dot d}+m \varepsilon_{\dot b\dot a}  \varepsilon_{\gamma(\alpha   }   \psi_{ \beta)  \dot d}-m\varepsilon_{\alpha\beta }    \psi_{\gamma( \dot a}   \varepsilon_{\dot b)\dot d }~.
\eea
Here we used \eqref{usefulFM} and \eqref{fmEoM}. Adding together these pieces, we obtain
\be
\left(\varepsilon_{\dot b\dot d }( \p_{\beta\gamma}+m\varepsilon_{\beta\gamma})   \psi_{\alpha \dot a}- \alpha   \dot a\leftrightarrow \beta\dot b\right)=\varepsilon_{\dot b\dot a}\p_{\alpha \beta }\psi_{\gamma\dot d}-2 m\varepsilon_{\alpha\beta}\psi_{\gamma( \dot a}   \varepsilon_{\dot b)\dot d }~.
\ee
Therefore, we find
\be
\left[\left\{Q_{\alpha a \dot a}, Q_{\beta  b \dot b}\right\},\psi_{\dot d}^\gamma\right]=i \varepsilon_{ab} \Big[\varepsilon_{\dot b\dot a}  \p_{\alpha \beta }   \psi_{\gamma  \dot d}-2 m\varepsilon_{\alpha\beta }    \psi_{\gamma( \dot a}   \varepsilon_{\dot b)\dot d }\Big]=-i \varepsilon_{  a   b}\varepsilon_{\dot a \dot b}\p_{\alpha\beta}\psi_{\dot d}^\gamma-2im\varepsilon_{ab}\varepsilon_{\alpha\beta }    \psi_{\gamma( \dot a}   \varepsilon_{\dot b)\dot d }~,
\ee  
which gives the desired result \eqref{SUSYAlgHyper}.

\subsec{3d $\CN=4$ Maxwell-Chern-Simons}\label{MCSexample}
In this section, we discuss the free $\CN=4$ $U(1)$ gauge theory and a version of the universal deformation in this QFT that gives rise to $\mathfrak{so}(4)_R$-preserving masses for all the degrees of freedom. This deformation is nothing but a CS term, which induces a mass for the gauge field and corresponding mass terms for its superpartners\footnote{It may be somewhat surprising that this procedure results in an $\CN=4$ theory. Unlike the examples considered in \cite{Gaiotto:2008sd}, here we consider a CS term in the presence of a standard gauge-kinetic term. Together, these two terms deform the SUSY algebra and hence allow us to be compatible with the discussion in footnote 4 of that paper. See also the related Lagrangian appearing in appendix E.1 of \cite{Lin:2005nh}.}
\begin{equation}\label{vectUnivApp}
\begin{aligned}
S &= g^{-2}\int d^3x\Bigg[   F^{\mu\nu}F_{\mu\nu} - \p^\mu \Phi^{\da\db} \p_\mu\Phi_{\da\db} + i\lambda^{\alpha a \db} (\gamma^\mu )_{\alpha}{}^\beta\p_\mu \lambda_{\beta a \db}- D^{ab}D_{ab} \\
&\qquad\qquad\qquad 
 -2i m\varepsilon_{\mu\nu\rho} F^{\mu\nu}A^\rho+2i m\lambda^{\beta a \db}\lambda_{\beta a \db} -4m^2 \Phi^{\da\db}\Phi_{\da\db}\Bigg]~
\\
&= \int d^3x\Bigg[g^{-2} \Big( F^{\mu\nu}F_{\mu\nu} - \p^\mu \Phi^{\da\db} \p_\mu\Phi_{\da\db} + i\lambda^{\alpha a \db} (\gamma^\mu )_{\alpha}{}^\beta\p_\mu \lambda_{\beta a \db}- D^{ab}D_{ab}\Big)\\
&\qquad\qquad\qquad +i {k\over4\pi} \varepsilon_{\mu\nu\rho} F^{\mu\nu}A^\rho - i {k\over4\pi} \lambda^{\beta a \db}\lambda_{\beta a \db} - {k^2g^2\over16\pi^2} \Phi^{\da\db}\Phi_{\da\db}\Bigg]~,\\
\end{aligned}
\end{equation} 
where $F_{\mu\nu}= \partial_{\mu}A_{\nu}-\partial_{\nu}A_{\mu} $, while the universal mass, $m$, and Chern-Simons level $k\in \bZ$ are related as follows
\begin{equation}
m=-{kg^2\over 8\pi}~.
\end{equation}

  One can check that \eqref{vectUnivApp} is invariant under the following transformations
\begin{eqnarray}
[Q_{\alpha c \dot{c}}, A_\mu] &=& \frac{i}{2} (\gamma_\mu)_\alpha^{\ \beta} \lambda_{\beta c \dot{c}}~,\cr
[Q_{\alpha c \dot{c}},\Phi_{\dot{a}\dot{b}}] &=& \frac{1}{2}(\varepsilon_{\dot{a}\dot{c}} \lambda_{\alpha c \dot{b}} + \varepsilon_{\dot{b}\dot{c}} \lambda_{\alpha c \dot{a}})~,\cr
[Q_{\beta f \dot{f}},\lambda_{\alpha a \dot{b}}]&=& \frac{i}{2} (\gamma_\rho)_{\alpha \beta}\, \varepsilon_{a f} \varepsilon_{\dot{b}\dot{f}}  \varepsilon^{\mu\nu\rho} F_{\mu\nu} + \varepsilon_{\alpha\beta}\varepsilon_{\dot{b}\dot{f}}  D_{a f} + i (\gamma^\rho)_{\alpha \beta}\, \varepsilon_{a f}\, \p_\rho \Phi_{\dot{f}\dot{b}}+2im(\varepsilon_{\alpha\beta} \varepsilon_{af} \Phi_{\dot{f}\db})~,\cr
[Q_{\alpha d \dot{d}},D_{c f}] &=& -\frac{i}{2}\p_{\mu}\Big(\varepsilon_{cd} (\gamma^{\mu})_\alpha^{\ \beta} \lambda_{\beta f \dot{d}} + \varepsilon_{fd} (\gamma^{\mu})_\alpha^{\ \beta} \lambda_{\beta c \dot{d}}\Big)-i m \Big(\varepsilon_{cd}\lambda_{\alpha f \dot{d}} + \varepsilon_{fd}\lambda_{\alpha c \dot{d}}\Big)~. \ \ \ 
\end{eqnarray}

Through some straightforward but tedious algebra, it is possible to show that 
\begin{eqnarray}\label{SUSYtransVecApp}
\left[\{Q_{\alpha a \dot{a}}, Q_{\beta b \dot{b}}\},A_{\mu}\right]&=&-i\varepsilon_{ab}\varepsilon_{\da\db}\partial_{\alpha\beta}A_{\mu}+\p_\mu \Lambda_{\alpha a \dot{a}, \beta b \dot{b}}~,\cr
\left[\{Q_{\alpha a \dot{a}}, Q_{\beta b \dot{b}}\},\Phi_{\dd\de}\right] &=&-i\varepsilon_{ab}\varepsilon_{\da\db}\partial_{\alpha\beta} \Phi_{\dd\de}+  mi \varepsilon_{\alpha\beta} \varepsilon_{ab}\left(\varepsilon_{\db\dd}\Phi_{\da\de} + \varepsilon_{\da\dd}\Phi_{\db\de}+ \varepsilon_{\db\de}\Phi_{\dd\da} +\varepsilon_{\da\de}\Phi_{\dd\db}\right)~,\cr
 \left[\{Q_{\alpha a \dot{a}}, Q_{\beta b \dot{b}}\},\lambda_{\kappa c \dot{c}}\right]&=&-i\varepsilon_{ab}\varepsilon_{\da\db}\partial_{\alpha\beta} \lambda_{\kappa c \dot{c}} +  mi \varepsilon_{\alpha\beta} \varepsilon_{ab}(\varepsilon_{\da\dc} \lambda_{\kappa c\db} + \varepsilon_{\db\dc} \lambda_{\kappa c\da})- mi \varepsilon_{\alpha\beta} \varepsilon_{\da \db} (\varepsilon_{ac} \lambda_{\kappa b\dc} + \varepsilon_{bc} \lambda_{\kappa a\dc})~,\cr
\left[\{Q_{\alpha a \dot{a}}, Q_{\beta b \dot{b}}\},D_{cf}\right]&=&-i\varepsilon_{ab}\varepsilon_{\da\db}\partial_{\alpha\beta}D_{cf}- mi \varepsilon_{\alpha\beta}\varepsilon_{\da\db}\left(\varepsilon_{bc}D_{af} + \varepsilon_{ac}D_{bf} + \varepsilon_{bf}D_{ca} + \varepsilon_{af}D_{cb}\right)~,
\end{eqnarray}
where $\Lambda_{\alpha a \da,\beta b \dot{b}} = -\varepsilon_{\alpha\beta}\varepsilon_{ab} \Phi_{\da\db}$ is a gauge transformation parameter. Since the non-trivial $R$-symmetries act as (we only show the non-trivial ones)
\begin{eqnarray}\label{RcommVec}
[(R_C)_{\da\db},\Phi_{\dd\de}]&=& i(\varepsilon_{\db\dd}\Phi_{\da\de} + \varepsilon_{\da\dd}\Phi_{\db\de}+ \varepsilon_{\db\de}\Phi_{\dd\da} +\varepsilon_{\da\de}\Phi_{\dd\db})~,\cr
[(R_C)_{\da\db},\lambda_{\kappa c\dc}]&=& i(\varepsilon_{\da\dc} \lambda_{\kappa c\db} + \varepsilon_{\db\dc} \lambda_{\kappa c\da})~,\cr
[(R_H)_{ab},\lambda_{\kappa c\dc}]&=& i(\varepsilon_{\da\dc} \lambda_{\kappa c\db} + \varepsilon_{\db\dc} \lambda_{\kappa c\da})~,\cr
[(R_H)_{ab},D_{cf}]&=& i(\varepsilon_{bc}D_{af} + \varepsilon_{ac}D_{bf} + \varepsilon_{bf}D_{ca} + \varepsilon_{af}D_{cb})~.
\end{eqnarray}
we see that this resulting supersymmetry algebra  is indeed given by  \eqref{defAlgApp}, up to a gauge transformation.  

More generally, we can imagine a $U(1)^N$ Maxwell-Chern-Simons theory (such theories are relevant in all the examples in the main text, since all of our theories either contain multiple $U(1)$ gauge groups or are mirror dual to theories that do) and their supersymmetric completions. In this case, we can generalize \eqref{vectUnivApp} as follows
\begin{eqnarray}\label{eq:quiverLag}
S &=& \int d^3x\Bigg[ \vec{F}^{\mu\nu}\cdot G^{-1} \cdot \vec{F}_{\mu\nu} - \p^\mu \vec{\Phi}^{\da\db} \cdot G^{-1} \cdot \p_\mu\vec{\Phi}_{\da\db} + i\vec{\lambda}^{\alpha a \db} \cdot G^{-1} \cdot (\gamma^\mu \p_\mu \vec{\lambda})_{\alpha a \db}- \vec{D}^{ab} \cdot G^{-1} \cdot \vec{D}_{ab}   \cr
& &\qquad \quad+  {i \over4\pi} \varepsilon_{\mu\nu\rho} \vec{F}^{\mu\nu} \cdot K \cdot \vec{A}^\rho -{i \over4\pi}\vec{\lambda}^{\beta b \db} \cdot K \cdot \vec{\lambda}_{\beta b \db} - \frac{1}{16 \pi^2} \vec{\Phi}^{\db\dc} \cdot K^2G^2 \cdot \vec{\Phi}_{\db\dc}\Bigg]~,
\end{eqnarray}
where each field has multiple components  forming a vector and the dot is just the matrix multiplication.  In \eqref{eq:quiverLag}, we have defined a symmetric matrix of inverse gauge couplings, $G^{-1}$, and a symmetric integral CS level matrix, $K$. Here we have the relation
\begin{equation}\label{eq:KGM}
K = -8\pi G^{-1}M~.
\end{equation} 
Note that, since the $K$ matrix is symmetric, we have
\begin{equation}
K^T =-8\pi M^T (G^{-1})^T =-8\pi M G^{-1} = K = -8\pi G^{-1}M\,.
\end{equation}
In other words, $M$ commutes with $G^{-1}$ and hence $G$.\footnote{As a result, all the matrices $M,K,G$ are diagonal in a proper basis.} Through a similar set of manipulations as in the $N=1$ case, we can show that the theory is invariant under the obvious generalization of \eqref{SUSYtransVecApp}. In this case, the SUSY algebra is
\begin{equation}
\{Q_{\alpha a \dot{a}}, Q_{\beta b \dot{b}}\}_{ij} = \varepsilon_{ab}\varepsilon_{\da\db} P_{\alpha\beta} \delta_{ij} + M_{ij}\Big( \varepsilon_{\alpha\beta} \varepsilon_{ab} (R_C)_{\da \db} - \varepsilon_{\alpha\beta} \varepsilon_{\da \db} (R_H)_{a b}\Big)~.
\end{equation}

\subsec{Interacting theories}\label{interactingApp}
At least classically, we can define interacting 3d $\CN=4$ theories invariant under \eqref{defAlgApp} that couple vector multiplets and hypermultiplets. The main point is that the universal mass parameters of the gauge and matter sectors should agree so that a common supersymmetry Algebra is preserved. 

For example, taking the vector multiplet action in \eqref{vectUnivApp}, we can couple these degrees of freedom to a hypermultiplet as follows
\begin{eqnarray}
\mathcal{L}_{\text{HM}} &=&D^\mu\tilde  \rho^a D_\mu  \rho_a - i \tpsi^{\da}  \gamma^\mu  D_\mu \psi_{\da} + i \tilde\rho^a D_a^{\ b} \rho_b - \frac{1}{2} \tilde\rho^a \Phi^{\da\db} \Phi_{\da\db}\rho_a - i\tilde \psi^{\da} \Phi_{\da}^{\ \db} \psi_{\db} + i \tilde\rho^a \lambda_a^{\ \db} \psi_{\db} + i \tpsi^{\da} \lambda^b_{\ \da} \rho_b\cr
& & +m^2 \tilde\rho^a \rho_a + i m \tpsi^{\da} \psi_{\da}~.
\end{eqnarray}
where $D_\mu$ are covariant derivatives.
Related Lagrangians have appeared in \cite{Lin:2005nh} (see also the discussion in \cite{Itzhaki:2005tu}).

\newpage

\end{appendices}

\newpage
\nocite{*}
\bibliography{chetdocbib}

\end{document}